# ♯SHAARP: An Open-Source Package for Analytical and Numerical Modeling of Optical Second Harmonic Generation in Anisotropic Crystals

Rui Zu,[1,a] Bo Wang[1,5,a], Jingyang He[1], Jian-Jun Wang[1], Lincoln Weber[1], Long-Qing Chen[1,3,4], Venkatraman Gopalan[1,2,3]

[a]These authors contribute equally to this work

[1]Department of Materials Science and Engineering, The Pennsylvania State University, University Park, Pennsylvania 16802, USA

[2] Department of Physics, Pennsylvania State University, University Park, Pennsylvania, 16802, USA

[3]Department of Engineering Science and Mechanics, The Pennsylvania State University, University Park, Pennsylvania 16802, USA

[4]Department of Mathematics, The Pennsylvania State University, University Park, Pennsylvania 16802, USA

[5]Materials Science Division, Lawrence Livermore National Laboratory, Livermore, CA 94550, USA

Venkatraman Gopalan (vxg8@psu.edu); Long-Qing Chen (lqc3@psu.edu); Rui Zu (ruizu0110@gmail.com); Bo Wang (bzw133.psu@gmail.com)

## Abstract

Optical second harmonic generation is a second-order nonlinear process that combines two photons of a given frequency into a third photon at twice the frequency. Due to the symmetry constraints, it is widely used as a sensitive probe to detect broken inversion symmetry and local polar order. Analytical modeling of the electric-dipole SHG response is essential to extract fundamental properties of materials from experiments. However, complexity builds up dramatically in the analytical model when the probed crystal is of a low bulk crystal symmetry, with a low-symmetry surface orientation, exhibits absorption and dispersion, and consists of multiple interfaces. As a result, there is a largely uneven landscape in the literature on the SHG modeling of new materials, involving numerous approximations and a wide range of (in)accuracies, leading to a rather scattered dataset of reported SHG nonlinear susceptibility. Towards streamlining the reliability and accuracy of this process, we have developed an open-source package called the Second Harmonic Analysis of Anisotropic Rotational Polarimetry (♯SHAARP) which derives analytical solutions and performs numerical simulations of reflection SHG from a single interface for homogeneous crystals. Five key generalizations in SHG modeling are implemented, including all crystal symmetries down to triclinic, any crystal orientation, complex dielectric tensor (refractive indices) with frequency dispersion, and general polarization states of the light. ♯SHAARP enables accurate anisotropic modeling of SHG response for a broad range of materials systems. The method is extendible to multiple interfaces. The code is free to download from https://github.com/Rui-Zu/SHAARP

Github: https://github.com/Rui-Zu/SHAARP

Documentation: https://shaarp.readthedocs.io/en/latest/

## Introduction

The ability to combine and split photons using nonlinear optical interactions has had a dramatic impact on generating a broad and continuously tunable electromagnetic spectrum towards furthering both fundamental science as well as technological applications. Extreme 100$^{th}$ harmonics are used to generate x-rays and deep ultraviolet for spectroscopy, diffraction, and medical imaging; the near-infrared (IR) laser light at 1.55$\mu$m powers the internet; the mid-and far-IR extending to the THz cover the fingerprint region for chemical sensing, environmental monitoring, free-space communication, gravitational wave detection, homeland security, aviation, medical imaging, and laser surgery.[1–5] We are presently at the threshold of a new era of quantum communications. Nonlinear optics remains the primary means to generate entangled photons today, with the promise to revolutionize secure communications, sensing, and metrology.[6]

Optical second harmonic generation (SHG) is a nonlinear optical process in a material that combines two incident photons at frequency $\omega$ into one photon at frequency $2\omega$, i.e. $\omega + \omega = 2\omega$ indicating energy conservation. It is a specific instance of three wave-mixing processes in which the frequency of each of the three photons could be different, i.e. $\omega_1 \pm \omega_2 = \omega_3$, where the positive sign corresponds to sum-frequency generation (SFG) and the negative sign to the difference frequency generation (DFG). More generally, higher-order nonlinear optical processes involving four or more waves are also possible.

Electric-dipole optical SHG, the focus of this work, is described by the interaction $\boldsymbol{P}^{2\omega} = \varepsilon_0 \boldsymbol{\chi}^{(2)} \boldsymbol{E}^{\omega} \boldsymbol{E}^{\omega}$, where $\boldsymbol{\chi}^{(2)}$ is the SHG nonlinear susceptibility tensor relating the fundamental electric field vectors $\boldsymbol{E}^{\omega}$ with frequency $\omega$ inside the crystal to the creation of a nonlinear polarization, $\boldsymbol{P}^{2\omega}$ at frequency $2\omega$. The electric-dipole SHG tensor, $\boldsymbol{\chi}^{(2)}$, is a third-rank polar tensor that keeps track of the polarizations of all three photons involved in the three-wave mixing process. As a consequence of Neumann's principle[7], it contains non-zero components only in materials that lack spatial inversion symmetry, or so-called non-centrosymmetric materials. Therefore, electric-dipole SHG is used as a sensitive probe of inversion symmetry breaking and polar order.

The classical, semi-classical, and quantum theories of nonlinear optical processes are well established.[8–10] Based on the fundamental theories, analytical and numerical approaches have been utilized to model the nonlinear optical responses. The analytical approach is essential to derive a closed-form expression of the SHG intensity that connects experimental observations to material fundamental properties. When material properties are already established, numerical simulations can be used to predict the SHG responses of samples with inhomogeneous microstructures and/or with complex measurement geometries in which an analytical solution is intractable. The numerical approach solves Maxwell equations with well-defined boundary conditions using techniques such as the finite element method and finite difference time domain method.[11]

Bloembergen, Maker, and later Herman & Hayden and Shoji et. al. laid out the analytical theory of SHG interactions at a single interface as well as in a slab geometry in both reflection and transmission geometries.[8–10,12,13] Two common models used to quantitively obtain absolute SHG coefficients are the Maker fringes[9] technique and the Bloembergen and Pershan formulae[8]. The Maker fringes can be obtained by measuring the transmitted SHG intensity as a function of the incident angle of a slab sample with both surfaces parallel to each other. The SHG coefficients of the slab can then be accurately obtained by analyzing the envelope of the curve.[10,14] Nevertheless, the Maker fringes technique is generally limited to the characterization of SHG properties of materials with small absorption, because of the transmission geometry. In contrast, other methods based on the reflection geometry provide greater flexibility, loosening the strict requirement for a transparent crystal, and allowing for the characterization of highly absorbing and reflecting bulk crystals such as metallic materials as well as ultrathin samples on an absorbing substrate such as few-layer flakes of 2D materials.[15,16]

However, most of these theoretical studies adopt some simplifications to keep the analytical solutions tractable. The probed crystals are generally limited to optically isotropic,[8] uniaxial,[10,14] or sometimes orthorhombic (biaxial) crystals[12,17] but cut along a high-symmetry surface to reduce complexity. The material is also often assumed to be transparent and non-absorptive at the pump and second harmonic frequencies.[8,10,17] A general analytical solution has not yet been established for modeling SHG polarimetry in nonlinear optical materials in all three optical classes (anaxial, uniaxial, biaxial)[18], including absorption and dispersion, and for any surface orientation geometry. ♯SHAARP addresses this critical need.

In contrast to analytical methods, there have been many well-established numerical approaches in the field of computational electromagnetics that can model nonlinear optical response[15,19,20]. Based on these numerical approaches, commercial and open-source software packages are available, including COMSOL Multiphysics[15], CST Studio Suite[19], and MEEP[20].

Complementary to these tools, ♯SHAARP has unique advantages in the following respects. First, ♯SHAARP can provide fully analytical or semi-analytical solutions to the Maxwell equations for the reflected or transmitted SHG waves at a single interface (*.si*). (A code for multiple interfaces, *.mi*, is under development). With these analytical expressions, users can fit their experimental measurements of the SHG polarimetry in a relatively straightforward manner. Particularly in this process, the users can determine whether to impose specific assumptions to simplify the final expression and evaluate quantitatively how each assumption can influence the results. It also standardizes the process and eliminates errors in publications, where each user does not have to derive specific (often messy) equations for their sample and their geometry. Second, since ♯SHAARP solves the Maxwell equations analytically, it can calculate the numerical results very efficiently. This feature allows one to predict the propagation directions and intensities of the SHG waves for a given system with known material properties in an on-demand modality without requiring running a numerical simulation for each desired change in variables. Third, ♯SHAARP is designed with a user-friendly GUI, which can guide the users to specify the inputs and conveniently export the output, including figure plots, numerical data, and analytical expressions. There is no need to build a finite-size system or specify the boundary conditions, and coding experience is not a prerequisite. Besides, ♯SHAARP is freely available to users via Mathematica® and Wolfram Player® and is expandable; while the former needs a license, the latter is free to use. The users may access the original code from GitHub and contribute to the code by optimizing it and adding new functionalities. We also plan to integrate new functionalities into ♯SHAARP in future versions, such as the SHG waves across slabs and multilayers. Based on these attributes, we believe ♯SHAARP can be as a critical missing piece in the fields of nonlinear optics and optical materials research.

In this work, an open-source package, ♯SHAARP (Second Harmonic Analysis of Anisotropic Rotational Polarimetry) is developed to calculate analytical and numerical solutions for the polarization-dependent SHG generated from a single interface in the frequency domain. The software is designed to study the SHG of an absorbing bulk crystal or a wedged slab sample from a single interface in reflection geometry. Five key attributes are integrated into ♯SHAARP: arbitrary symmetry, arbitrary crystal orientation, arbitrary incident and measured polarizations of light, and the inclusion of dispersion and absorption. To benchmark the theoretical calculations, the reflective SHG intensity of three commercial crystals, namely, GaAs, $LiNbO_3$, and $KTiOPO_4$ (KTP), corresponding to the isotropic (anaxial), uniaxial, and biaxial classes, respectively, are experimentally measured and fitted using the analytical theory provided by ♯SHAARP to extract their nonlinear tensors and validate against published literature. Excellent agreement between this work and literature demonstrates a robust means of understanding and characterizing anisotropic second harmonic response and nonlinear coefficients. An example is also given where ♯SHAARP aided analysis of a new material, TaAs, as compared to the published literature.[21,22]

## SHG Polarimetry

SHG measurement is often performed using SHG polarimetry in various geometries as shown in **Fig. 1**. Polarimetry refers to the mapping of the SHG intensities for various combinations of the polarization of the incident fundamental photon at $\omega$ and the output SHG photon at $2\omega$. This can occur in transmission or reflection geometries, and the polarization of the incident and SHG light can each be, in general, elliptical with the special cases of linear and circular polarizations; these polarization states can further be either rotated or fixed. **Figures 1(a) and 1(d)** depict the schematics of two common experimental geometries where the red and blue rays are the fundamental and second harmonic light, respectively. Here, we will refer to **Figure 1(a)** as the rotating polarizer, fixed analyzer (FA) geometry, and **Figure 1(d)** as the rotating polarizer, rotating analyzer (RA) geometry, for the following discussion. The FA geometry in panel (**a**) is commonly achieved using a rotating half-wave plate to rotate the linear polarization of the incident fundamental wave while using a fixed analyzer in the *s* or *p* polarization geometries for the second harmonic wave. The RA geometry in panel (**d**) is achieved by either fixing both the half-wave plate and the analyzer while rotating the crystal or by rotating the half-wave plate and the analyzer simultaneously while fixing the sample orientation. The plane of incidence (PoI) is defined as the $L_1 - L_3$ plane, where $(L_1, L_2, L_3)$ are the lab coordinates as displayed in **Fig. 1a** and **d**. In both geometries, the incident linear polarization ($\boldsymbol{E}^{i,\omega}$) is rotated, while second harmonic intensities are collected as a function of the azimuthal angle $\varphi$ in two different ways. Here, superscripts $i$ and $\omega$ respectively represent incident light and fundamental frequency. $\varphi = 0$ represents the *p*- polarized $\omega$ wave in the FA geometry, and the *p*- polarized $\omega, 2\omega$ waves in the RA geometry. In the FA geometry of **Fig. 1a**, the *p* and *s* polarized second harmonic intensities ($I_p^{2\omega}(\varphi)$ and $I_s^{2\omega}(\varphi)$) are measured as depicted in panels (**b**) and (**c**). In the RA geometry of **Fig. 1d**, the second harmonic intensities polarized either parallel or perpendicular to the incident fundamental polarization, namely, $I_{\parallel}^{2\omega}(\varphi)$ and $I_{\perp}^{2\omega}(\varphi)$ are measured as depicted in panels (**e**) and (**f**). **Figures 1(c)** and **(f)** depict the calculated SHG polar plots of GaAs (111) obtained in the normal incidence geometry, which contain information about crystal symmetry, linear dielectric, or refractive index tensors at both $\omega$ and $2\omega$ frequencies, and second-order nonlinear susceptibility tensor.

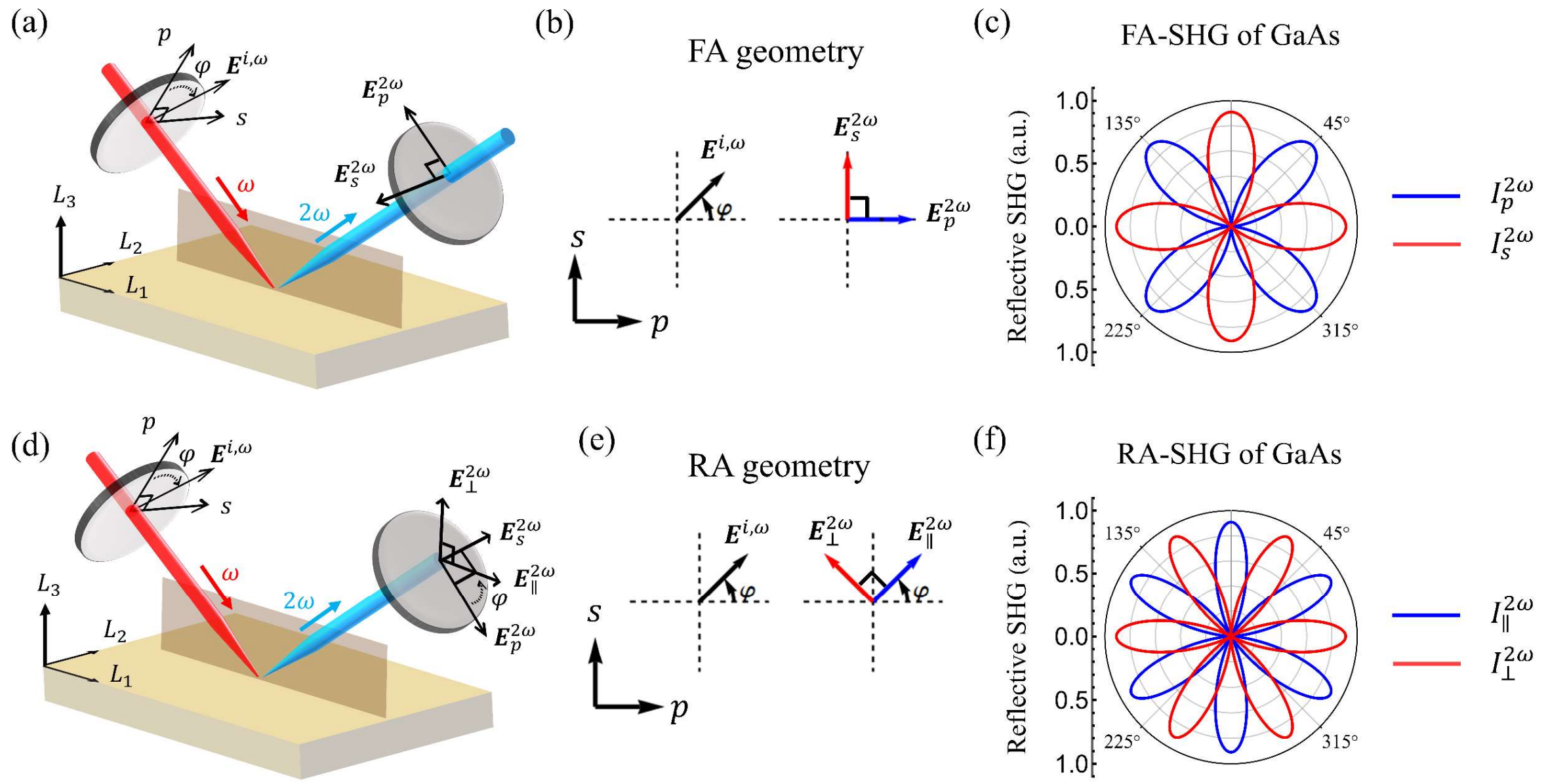

Figure 1. Two common experimental geometries for SHG polarimetry and the resulting reflection SHG polarimetry for GaAs(111) surface in normal incidence. **a-c** Rotating polarizer fixed analyzer (FA) geometry: SHG intensities of $p$- and $s$- polarized waves are recorded as a function of the azimuthal angle $\varphi$. **d-f** Rotating polarizer, rotating analyzer (RA) geometry: SHG intensities polarized parallel or perpendicular to the incident polarization are recorded as a function of the azimuthal angle $\varphi$. **a,d** Schematics of the experimental geometries is shown. $(L_1, L_2, L_3)$ is the lab coordinate system. The sample surface is located in the $L_1 - L_2$ plane, while the plane of incidence is parallel to the $L_1 - L_3$ plane. Red and blue waves indicate a pump beam at $\omega$ and signal beam at $2\omega$ frequency, respectively. **b,e** Schematics showing the relation between the incident polarization and SHG polarization projected in the $E_p - E_s$ plane for the two experimental geometries. **c,f** Polar plots of the calculated reflective SHG intensities for GaAs(111) surface subject to normal incidence using the two geometries.

## Theoretical background

Consider an incident plane wave with the fundamental frequency $\omega$ onto a flat surface of a noncentrosymmetric crystal. The incident wave results in the generation of transmitted and reflective waves with $\omega$ and $2\omega$ frequencies, as schematically illustrated in **Fig. 2a**. Each wave can be denoted by its wavevector $\boldsymbol{k}$, the superscript of which indicates the frequency ($\omega$ or $2\omega$) as well as the birefringence. Here, $e$ and $o$ are for the two Eigen waves at each frequency, while $ee$, $oo$, and $eo$ are for the nonlinear $2\omega$ waves generated from three distinct combinations of the fundamental ($\omega$) $e$ and $o$ waves. Note that the $s$ and $p$ polarization states are, in general, distinct from the $e$ and $o$ polarization states. The $s$ and $p$ polarization states are defined as the electric field ($\boldsymbol{E}$) being perpendicular ($s$) or parallel ($p$) to the PoI (the plane defined by the incidence wavevector and the sample surface normal). The $e$ and $o$ polarization states are defined as the dielectric displacement ($\boldsymbol{D}$) with a component along ($e$) or perpendicular to ($o$) the optical axes (0 optical axis for anaxial, 1 for uniaxial, and 2 for biaxial crystals). The extraordinary wave ($e$) is normal to $\boldsymbol{D}^o$ (ordinary polarized dielectric displacement).[23] The other superscripts, $i$, $R$, or $T$, correspond to the incident, reflected, or transmitted beams, respectively. The angle $\theta$ represents the propagating angle of the associated wave with the wave vector $\boldsymbol{k}$.

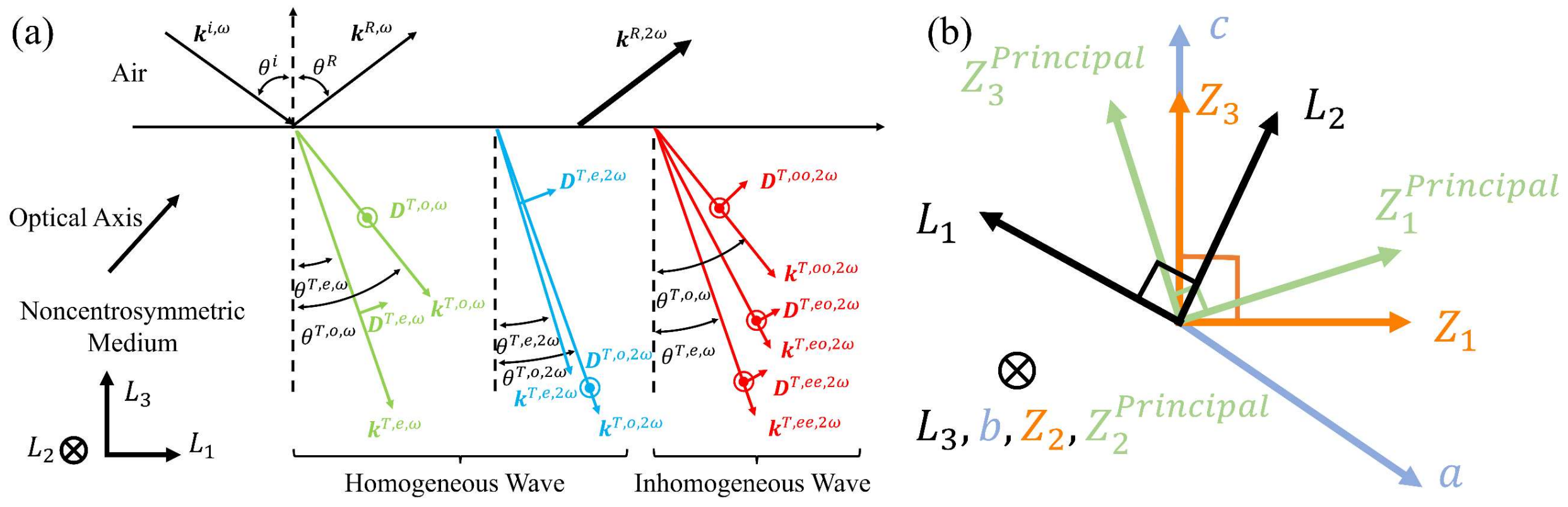

Figure 2. **a** Schematic example of different waves at both $\omega$ and $2\omega$ frequencies for the specific measurement geometry where the optic axes are chosen to lie in the incidence plane for clarity sake. One optical axis is shown and the other one will be in the plane as well if it is biaxial material. ♯SHAARP does not in general have such a restriction, and can handle an arbitrary orientation of the optic axes with respect to the incidence plane. The green and blue rays are homogeneous waves at $\omega$ and $2\omega$ frequency. The red rays are the inhomogeneous waves at $2\omega$. **b** Schematic of four different coordinate systems used for a monoclinic crystal structure. $(L_1, L_2, L_3)$, ($a$, $b$, $c$), $(Z_1, Z_2, Z_3)$, and $(Z_1^{Principal}, Z_2^{Principal}, Z_3^{Principal})$ are the lab, crystallographic, crystal physics, and principal coordinate systems, respectively. Only the crystallographic coordinate system is non-orthogonal.

Four different coordinate systems shown in **Fig. 2b** are necessary to describe the SHG measurement, and it is essential to clarify their mutual relationships. $(L_1, L_2, L_3)$ describe the *lab coordinate system* (LCS), where $L_3$ is defined as the normal to the sample surface, $L_2$ is selected as being perpendicular to the PoI, and $L_1$ is taken to the direction that ensures this coordinate system is orthogonal. For a crystal under study, the translation vectors of the unit cell of the crystal determine the crystallographic coordinate system (CCS) given by ($a$, $b$, $c$); these axes may *not* be orthogonal. The $(Z_1, Z_2, Z_3)$ represent the *crystal physics coordinate system* (ZCS) in which the material property tensors are represented; they are always orthogonal and their directions relative to ($a$, $b$, $c$) follow the IEEE standards as summarized in **Table. S2**.[24,25] The $(Z_1^{Principal}, Z_2^{Principal}, Z_3^{Principal})$ is the *principal coordinate system* (PCS), in which the dielectric tensor, and hence the refractive index ellipsoid are diagonalized; this coordinate system is also always orthogonal. For an isotropic or uniaxial crystal structure, $(Z_1, Z_2, Z_3) = (Z_1^{Principal}, Z_2^{Principal}, Z_3^{Principal})$. For a biaxial crystal, the PCS is *defined* with the real part of the refractive indices along each axis following an ascending order[23] as follows: $n(Z_1^{Principal}) < n(Z_2^{Principal}) < n(Z_3^{Principal})$, while this is *not* true in general for refractive indices defined in the crystal physics coordinate system. Henceforth, we will adopt the notation $\tilde{n}(Z_i^{Principal}) \equiv \tilde{n}_i^{\omega}$ for the eigenvalues of the refractive index tensor. In the PCS, the diagonal components of the *relative* dielectric tensor at optical frequency can be conveniently written as $\tilde{\boldsymbol{\varepsilon}}^{Principal} = \varepsilon_0(\tilde{\boldsymbol{n}})^2$, where $\varepsilon_0$ is the vacuum permittivity. For a material with complex dielectric tensors and refractive indices, the complex quantities can be expressed as $\tilde{\boldsymbol{\varepsilon}} = \boldsymbol{\varepsilon}_R + j\,\boldsymbol{\varepsilon}_I$ and $\tilde{\boldsymbol{n}} = \boldsymbol{n}_R + j\,\boldsymbol{n}_I$. Subscripts $R$ and $I$ represent real and imaginary components of the tensors, and $j = \sqrt{-1}$. Therefore, the dielectric permittivity $\tilde{\boldsymbol{\varepsilon}}^{\omega}$ in the LCS can be expressed as

$$\tilde{\boldsymbol{\varepsilon}}^{\omega} = \boldsymbol{a}_{LZ}\boldsymbol{a}_{ZP}\begin{pmatrix} \tilde{n}_1^{\omega} & 0 & 0 \\ 0 & \tilde{n}_2^{\omega} & 0 \\ 0 & 0 & \tilde{n}_3^{\omega} \end{pmatrix}^2 (\boldsymbol{a}_{LZ}\boldsymbol{a}_{ZP})^{-1} \tag{1}$$

where $\boldsymbol{a}_{LZ}$ is the rotation matrix from ZCS to the LCS, and $\boldsymbol{a}_{ZP}$ is the rotation matrix from the PCS to the ZCS.

When a monochromatic plane wave at frequency $\omega$ is incident upon the interface, two refracted rays at frequency $\omega$ can propagate inside the medium with one of two possible orthogonal dielectric displacement vectors $\boldsymbol{D}^{T,e,\omega}$ and $\boldsymbol{D}^{T,o,\omega}$. The two waves can be both ordinary, or one ordinary and one extraordinary, depending on the optical class of the material and the propagation direction of light. Without loss of generality, we denote the two transmitted (and subsequently refracted) waves at the single interface by superscript *T*, as shown in green in **Fig. 2a**. The two fundamental waves correspond to the Eigen solutions of the wave equation at the linear frequency, $\omega$ , given in the LCS as

$$\nabla \times \nabla \times \boldsymbol{E}^{T,\omega} + \begin{pmatrix} \varepsilon_{L_1L_1}^{\omega} & \varepsilon_{L_1L_2}^{\omega} & \varepsilon_{L_1L_3}^{\omega} \\ \varepsilon_{L_2L_1}^{\omega} & \varepsilon_{L_2L_2}^{\omega} & \varepsilon_{L_2L_3}^{\omega} \\ \varepsilon_{L_3L_1}^{\omega} & \varepsilon_{L_3L_2}^{\omega} & \varepsilon_{L_3L_3}^{\omega} \end{pmatrix} \mu^{\omega} \frac{\partial^2}{\partial t^2} \boldsymbol{E}^{T,\omega} = 0 \tag{2}$$

where $\varepsilon_{L_i,L_j}^{\omega}$ is the dielectric permittivity tensor at frequency $\omega$ in the LCS, and the $\mu^{\omega}$ represents the magnetic permeability at $\omega$.[23] Typically, for a non-magnetic medium, $\mu^{\omega} \sim \mu_0$, the vacuum permeability is assumed. In general, the anisotropic dielectric permittivity and magnetic permeability tensors of the medium are not diagonalized in LCS. Therefore, the non-collinearity between $\boldsymbol{E}$ and $\boldsymbol{D}$, as well as $\boldsymbol{B}$ and $\boldsymbol{H}$ results in two separate orthogonal bases, namely, ($\boldsymbol{k}$, $\boldsymbol{D}$, $\boldsymbol{B}$) and ($\boldsymbol{S}$, $\boldsymbol{E}$, $\boldsymbol{H}$). Here, $\boldsymbol{k}$, $\boldsymbol{D}$, $\boldsymbol{B}$, $\boldsymbol{S}$, $\boldsymbol{E}$, and $\boldsymbol{H}$ are respectively, the wavevector, dielectric displacement, magnetic induction, Poynting vector, electric field and magnetic field intensity.[26] Note that $\boldsymbol{E}$ and $\boldsymbol{H}$ are not necessarily orthogonal to the wavevector $\boldsymbol{k}$ inside the medium.

The nonlinear polarization induced by second-order nonlinear susceptibility radiates nonlinear source waves at $2\omega$ frequency, which can be written as,

$$\boldsymbol{P}^{2\omega} = \varepsilon_0 \boldsymbol{\chi}^{(2)} \boldsymbol{E}^{T,\omega} \boldsymbol{E}^{T,\omega} \exp i(\boldsymbol{k}^S \cdot \boldsymbol{r} - \omega t) \tag{3}$$

where $\boldsymbol{P}^{2\omega}$, $\boldsymbol{E}^{T,\omega}$, $\boldsymbol{\chi}^{(2)}$ are the nonlinear polarization, the electric field of the refracted $\omega$ waves, and the second-order nonlinear susceptibility. The term $\boldsymbol{k}^S$ is the wavevector of the source wave (superscript *S*) that combines two linear wavevectors[8], i.e., $\boldsymbol{k}^S = 2\boldsymbol{k}^{T,e,\omega}$, $2\boldsymbol{k}^{T,o,\omega}$ or $\boldsymbol{k}^{T,e,\omega} + \boldsymbol{k}^{T,o,\omega}$. The electric fields radiated by the nonlinear polarization can then be calculated in the LCS using the wave equation[8,27,28],

$$\nabla \times \nabla \times \boldsymbol{E}^{T,2\omega} + \begin{pmatrix} \varepsilon_{L_1L_1}^{2\omega} & \varepsilon_{L_1L_2}^{2\omega} & \varepsilon_{L_1L_3}^{2\omega} \\ \varepsilon_{L_2L_1}^{2\omega} & \varepsilon_{L_2L_2}^{2\omega} & \varepsilon_{L_2L_3}^{2\omega} \\ \varepsilon_{L_3L_1}^{2\omega} & \varepsilon_{L_3L_2}^{2\omega} & \varepsilon_{L_3L_3}^{2\omega} \end{pmatrix} \mu^{2\omega} \frac{\partial^2}{\partial t^2} \boldsymbol{E}^{T,2\omega} = -\mu^{2\omega} \frac{\partial^2}{\partial t^2} \boldsymbol{P}^{2\omega} \tag{4}$$

where $\varepsilon_{L_i,L_j}^{2\omega}$ represents the dielectric permittivity tensor at frequency $2\omega$ in the LCS, and the $\mu^{2\omega}$ represents the magnetic permeability at $2\omega$.[23] The homogeneous wave and inhomogeneous waves radiated by the nonlinear polarization correspond to the general and particular solutions of **Eq. (4)**, respectively. The former component is also known as the "free wave", and the latter is the radiated wave by the nonlinear polarization known as the "bound wave."[10,14] The total nonlinear wave can be expressed as a superposition of the general and particular solutions. To solve for the homogeneous wave at both linear (**Eq. 2**) and nonlinear (**Eq. 4**) frequencies, the *congruence transformation,* and *generalized Eigen equation* are employed.[29] The two eigenvalues correspond to the effective refractive indices and the two eigenvectors correspond to the electric field directions for the two homogeneous $e$ and $o$ waves. Three inhomogeneous waves $(\boldsymbol{k}^{T,ee,2\omega}, \boldsymbol{k}^{T,oo,2\omega}, \boldsymbol{k}^{T,eo,2\omega}) = (2\boldsymbol{k}^{T,e,\omega}, 2\boldsymbol{k}^{T,o,\omega}, \boldsymbol{k}^{T,e,\omega} + \boldsymbol{k}^{T,o,\omega})$, will be generated according to **Eq. 3**, as shown in **Fig. 2a**, due to the three-wave mixing process. Therefore, in principle, five waves at $2\omega$ will be generated, as shown in **Fig. 2a**, where blue and red correspond to homogeneous and inhomogeneous waves, respectively. The inhomogeneous SHG fields can be written in the following form:

$$\boldsymbol{E}^{T,ee,2\omega} = \boldsymbol{C}^{T,ee,2\omega} \exp i(\boldsymbol{k}^{T,ee,2\omega}.\boldsymbol{r} - \omega t)$$

$$\boldsymbol{E}^{T,oo,2\omega} = \boldsymbol{C}^{T,oo,2\omega} \exp i(\boldsymbol{k}^{T,oo,2\omega}.\boldsymbol{r} - \omega t)$$

$$\boldsymbol{E}^{T,eo,2\omega} = \boldsymbol{C}^{T,eo,2\omega} \exp i(\boldsymbol{k}^{T,eo,2\omega}.\boldsymbol{r} - \omega t) \tag{5}$$

where **C** is the field strength to be determined from **Equation (4)** for a given $\boldsymbol{P}^{2\omega}$. By substituting **Eq. (3)** and **(5)** into **Eq. (4)**, the field strengths of the three inhomogeneous waves can be explicitly calculated with the associated second-order optical susceptibilities. Accordingly, the $\boldsymbol{H}^{2\omega}$ for the three inhomogeneous waves can be obtained by

$$\boldsymbol{H}^{2\omega} = \frac{1}{\omega\mu_0}\boldsymbol{\mu}^{-1}\boldsymbol{k}^{2\omega} \times \boldsymbol{E}^{2\omega} \tag{6}$$

Boundary conditions at the interface (here, the surface of the crystal) are important to accurately determine the propagation directions and the field strengths of waves.[28] To satisfy the momentum conservation at both $\omega$ and $2\omega$ frequencies, it is required that

$$k_{L_1}^{i,\omega} = k_{L_1}^{R,\omega} = k_{L_1}^{T,e,\omega} = k_{L_1}^{T,o,\omega} \tag{7}$$

$$k_{L_1}^{R,2\omega} = k_{L_1}^{S,2\omega} = k_{L_1}^{T,e,2\omega} = k_{L_1}^{T,o,2\omega} \tag{8}$$

from which the wavevectors and propagation angles of all refractive and reflective waves can be determined. The continuity across the interface of the components, $E_{\parallel}$ and $H_{\parallel}$, of the electric and magnetic fields parallel to the interface, respectively, also yields the boundary conditions for the $\omega$ and $2\omega$ waves, given by

$$E_{\parallel}^{i,\omega} + E_{\parallel}^{R,\omega} = E_{\parallel}^{T,e,\omega} + E_{\parallel}^{T,o,\omega} = E_{\parallel}^{T,\omega} \tag{9}$$

$$H_{\parallel}^{i,\omega} + H_{\parallel}^{R,\omega} = H_{\parallel}^{T,e,\omega} + H_{\parallel}^{T,o,\omega} = H_{\parallel}^{T,\omega} \tag{10}$$

$$E_{\parallel}^{R,2\omega} = E_{\parallel}^{T,e,2\omega} + E_{\parallel}^{T,o,2\omega} + E_{\parallel}^{T,ee,2\omega} + E_{\parallel}^{T,oo,2\omega} + E_{\parallel}^{T,eo,2\omega} = E_{\parallel}^{T,2\omega} \tag{11}$$

$$H_{\parallel}^{R,2\omega} = H_{\parallel}^{T,e,2\omega} + H_{\parallel}^{T,o,2\omega} + H_{\parallel}^{T,ee,2\omega} + H_{\parallel}^{T,oo,2\omega} + H_{\parallel}^{T,eo,2\omega} = H_{\parallel}^{T,2\omega} \tag{12}$$

where superscripts $e$ and $o$ represent homogeneous waves and $ee$, $oo$ and $eo$ are inhomogeneous waves at $2\omega$ due to wave mixing. Using **Eq. (7) - (12)**, $\boldsymbol{E}^{R,2\omega}$ and $\boldsymbol{E}^{T,2\omega}$ can be calculated.

## Outline of the ♯SHAARP Code

We use a flowchart to illustrate the procedure for solving the equations using the boundary condition method for the $\omega$ and $2\omega$ waves in **Fig. 3**. All the relevant input and output variables in ♯SHAARP are summarized in **Table. S2**. The two refracted linear waves are described by solving **Eq. (2)**. Their actual field strengths can be obtained from the boundary conditions at the fundamental frequency $\omega$, using **Eqs. (7, 9, 10)**. Given **Eqs. 3** and **4**, two homogeneous waves and three inhomogeneous waves at $2\omega$ can be uniquely determined. Following the boundary condition analysis at $2\omega$ using **Eqs. (8, 11, 12)**, analytical expressions for the second harmonic response can be derived and used for simulations of polarimetry as well as for fitting experimental polar plots and extracting intrinsic SHG tensor coefficients.

Notably, we have made no assumptions above on the crystal symmetry, surface orientation, or absorptive nature of the material in deriving **Equations** (1) to (12). Therefore, the procedure is generally applicable to material systems with arbitrary crystal orientation, dielectric permittivity tensor, and SHG tensor. This general routine enables us to predict the SHG responses for given linear and nonlinear optical properties and to determine nonlinear optical coefficients by fitting experimental SHG polarimetry measurements of new materials with arbitrary crystal symmetry, surface orientation, and absorptive properties.

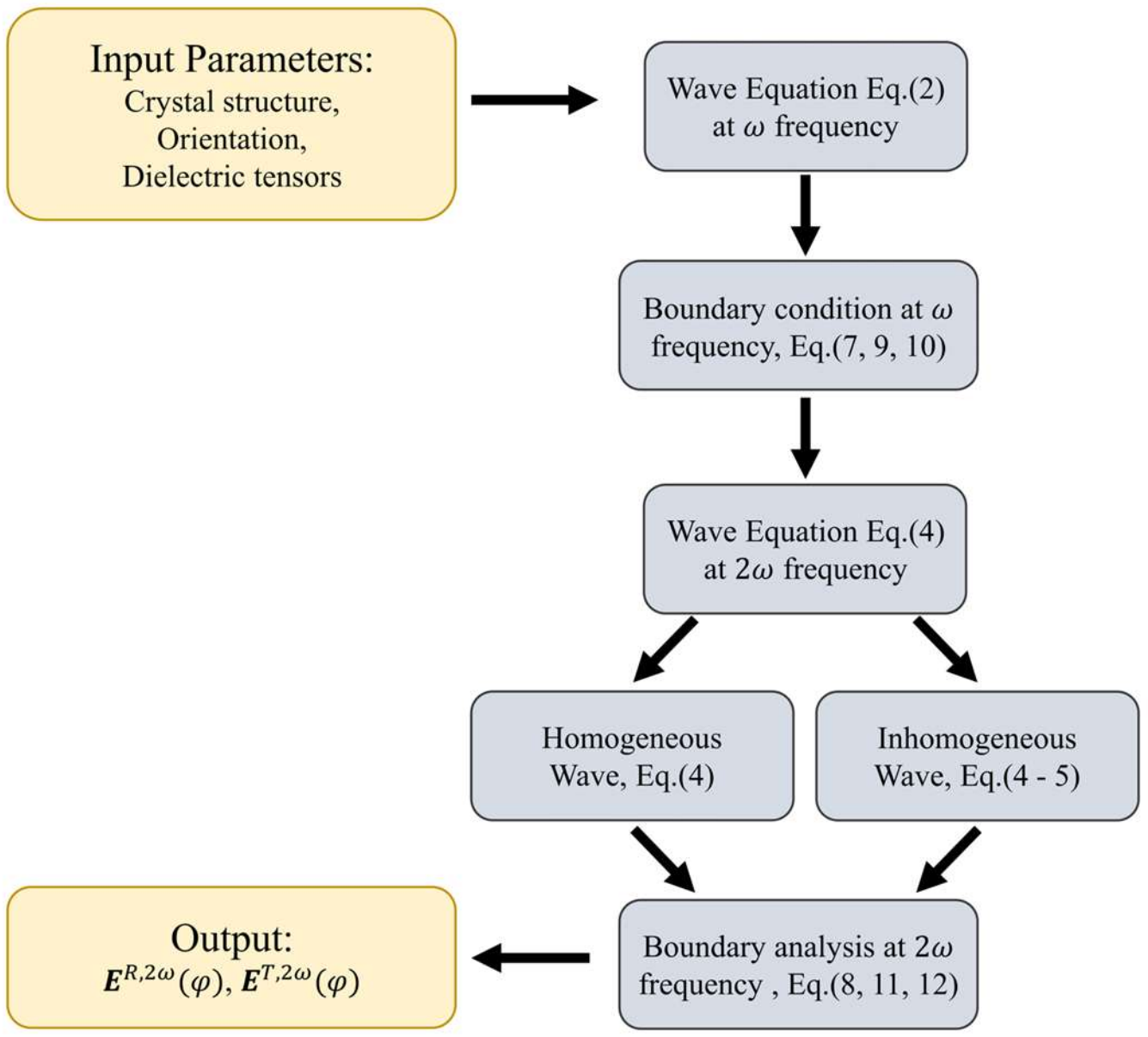

Figure 3. Flowchart illustrating the key steps in deriving the polarized second harmonic fields generated by a nonlinear medium.

To benchmark the theoretical results predicted by ♯SHAARP, we carried out SHG experiments on three typical nonlinear optical materials based on the FA geometry (c.f. **Fig. 1a**). To verify the capability of ♯SHAARP for materials of arbitrary symmetry, we chose three materials, GaAs, $LiNbO_3$, and KTP, which corresponds to three different optical classes: isotropic (also called anaxial, or lacking any optic axes), uniaxial (with one unique optic axis), and biaxial (with two unique optic axes). We also studied (112)-cut TaAs where the surface is parallel to a non-trivial crystallographic plane. Notably, among the materials we have chosen, the semiconducting GaAs is known for exhibiting a finite absorption of the fundamental and SHG waves, which allows us to test the capability of ♯SHAARP in modeling materials systems with absorption.

## Isotropic GaAs (111)

GaAs crystallizes in a cubic structure with the point group $\bar{4}3m$.[30] The CCS, ZPS, and PCS coincide and point in the same directions. It is one of the most widely used semiconductors, with a direct bandgap of around 1.42 eV [31–33], and can be patterned for quasi-phase-matching for nonlinear optical applications[34]. Using the fundamental probing energy of 1.55eV (as we have in this study), GaAs shows a finite absorption at both fundamental and second harmonic frequencies and exhibits strong resonances at $2\omega$ frequency.[31] GaAs (111) single crystal is used for the study, and it is oriented such that the $[1\bar{1}0]$ is parallel to the $L_2$ direction as shown in **Fig. 4a**. The dark yellow line corresponds to the projection of the PoI ($L_1 - L_3$ plane) on to the viewing plane $L_1 - L_2$.

At both the fundamental and second harmonic frequencies, the diagonalized complex dielectric tensors have three identical components due to the isotropic symmetry, i.e., $\tilde{\varepsilon}_{11} = \tilde{\varepsilon}_{22} = \tilde{\varepsilon}_{33}$. Consequently, the effective refractive index is independent of the incidence angle ($\theta^i$) as shown in **Fig. 4b**. The symmetry requires GaAs to possess only one independent nonzero second harmonic tensor component, i.e., $d_{14} = d_{25} = d_{36}$ , while the other components vanish.[7]

The intensities of the $p$- and $s$- polarized SHG waves, $I_s^{2\omega}(\varphi)$ and $I_p^{2\omega}(\varphi)$, are recorded at four different incident angles ($\theta^i = 0^\circ$, $15^\circ$, $30^\circ$ and $45^\circ$). For non-normal incidence, each polar plot contains at least three independent equations, namely $\cos(4\varphi)$, $\sin(4\varphi)$ and $\sin(\varphi)\cos(\varphi)$. Two independent equations at normal incidence are $\sin^2(2\varphi)$ and $\mathrm{Cos}^2(2\varphi)$. Thus, we used twenty equations to obtain the unique fitting of five unknown parameters (one SHG susceptibility and four geometric factors). The data are fitted to extract the nonlinear susceptibilities of GaAs. As shown in **Fig. 4c-d**, the open circles, and solid curves represent experimental results and the theoretical fittings, respectively, which demonstrate quantitatively good agreements. The analytical expressions for the fittings were generated by ♯SHAARP for the $\bar{4}3m$ point group with the probing geometry described above. The analytical solutions of reflected $2\omega$ waves for this most simplified case are given below,

$$E_p^{2\omega} = \frac{\left(C_{L_1}^{T,oo,2\omega} + C_{L_1}^{T,ee,2\omega} + C_{L_1}^{T,eo,2\omega}\right)\left(E_{L_1}^{T,e,2\omega}\tilde{n}^{2\omega}\cos\theta^{T,2\omega} - E_{L_1}^{T,e,2\omega}\tilde{n}^{\omega}\cos\theta^{T,\omega} + E_{L_3}^{T,e,2\omega}\tilde{n}^{2\omega}\sin\theta^{T,2\omega}\right) - \left(C_{L_3}^{T,oo,2\omega} + C_{L_3}^{T,ee,2\omega} + C_{L_3}^{T,eo,2\omega}\right)E_{L_1}^{T,e,2\omega}\tilde{n}^{\omega}\sin\theta^{T,\omega}}{E_{L_1}^{T,e,2\omega} + \tilde{n}^{2\omega}\cos\theta^{i}\left(E_{L_1}^{T,e,2\omega}\cos\theta^{T,2\omega} + E_{L_3}^{T,e,2\omega}\sin\theta^{T,2\omega}\right)} \quad (13)$$

$$E_s^{2\omega} = \frac{\left(C_{L_2}^{T,oo,2\omega} + C_{L_2}^{T,ee,2\omega} + C_{L_2}^{T,eo,2\omega}\right)\left(\tilde{n}^{2\omega}\cos\theta^{T,2\omega} - \tilde{n}^{\omega}\cos\theta^{T,\omega}\right)}{\tilde{n}^{2\omega}\left(\cos\theta^{T,2\omega} + \cos\theta^{i}\right)} \quad (14)$$

where $\left(E_{L_1}^{T,e,2\omega}, E_{L_2}^{T,e,2\omega}, E_{L_3}^{T,e,2\omega}\right) = \left(\cos\theta^{T,2\omega}, 0, \sin\theta^{T,2\omega}\right)$, is a unit vector describing the $p$-polarized electric field direction of one of the homogeneous waves at the $2\omega$ frequency. The variables, $C^{T,ee,2\omega}$, $C^{T,oo,2\omega}$, $C^{T,eo,2\omega}$ represent the amplitudes of the inhomogeneous waves radiated by the nonlinear polarization, and their explicit expressions are given in the supplementary section **S5**. Here, $\tilde{n}$ represents the complex refractive index. Subscripts $L_1$, $L_2$ and $L_3$ represent vector components along the lab coordinates, and $\theta^T$ and $\theta^i$ are the angles of refraction and incidence, respectively.

By fitting the data with expressions generated by ♯SHAARP, we have achieved excellent fitting between the analytical theory and the polarimetry experiments for all incident angles $\theta^i$, which confirms the $\bar{4}3m$ point group of GaAs. The absolute SHG coefficients are further obtained by calibrating the SH intensities against a reference nonlinear optical crystal ($LiNbO_3$ in this study) under the same probing conditions. The extracted absolute SHG coefficient, $d_{36}$ , at 800 nm is found to be $267 \pm 20$ pm/V, which agrees with the reported value, $310 \pm 50$ pm/V [35].

In many previous studies, various assumptions were made to simplify the analytical expressions in order to fit the experiments and to extract the absolute values of the SHG coefficients. To evaluate the influence of these assumptions on the calculated SHG tensor coefficients, we obtained the corresponding analytical expressions under several assumptions using ♯SHAARP and evaluate the SHG intensities accordingly, as shown in **Figure 4e**. The case labeled "♯SHAARP" represents our results based on **Eqs. 13** and **14** as described above. The results labeled by $\tilde{\boldsymbol{\varepsilon}}_R$ assumes negligible extinction coefficient, $\tilde{\boldsymbol{n}}_I$, and only the real part of the dielectric permittivity tensor (i.e. real refractive indices, $\tilde{\boldsymbol{n}}_R$) is used for deriving the analytical expression. Such assumptions are commonly employed in analyzing 2D materials with photon energy greater than their bandgaps, resulting in a largely uneven landscape of reported nonlinear susceptibilities.[36,37] The results labeled by $|\tilde{\boldsymbol{\varepsilon}}|$ takes the magnitude of relative dielectric constants instead of the complex quantities into the analysis, which folds in the effect of the absorption indirectly. One can see that, for the specific case of GaAs (111) considered, the second-order nonlinear susceptibility tends to be underestimated by nearly 20% if the extinction coefficient is neglected. In contrast, no significant change in the SH intensity is observed in this case if only the magnitude of the complex dielectric constants ($|\tilde{\boldsymbol{\varepsilon}}|$) instead of the actual complex dielectric tensor is taken into account.

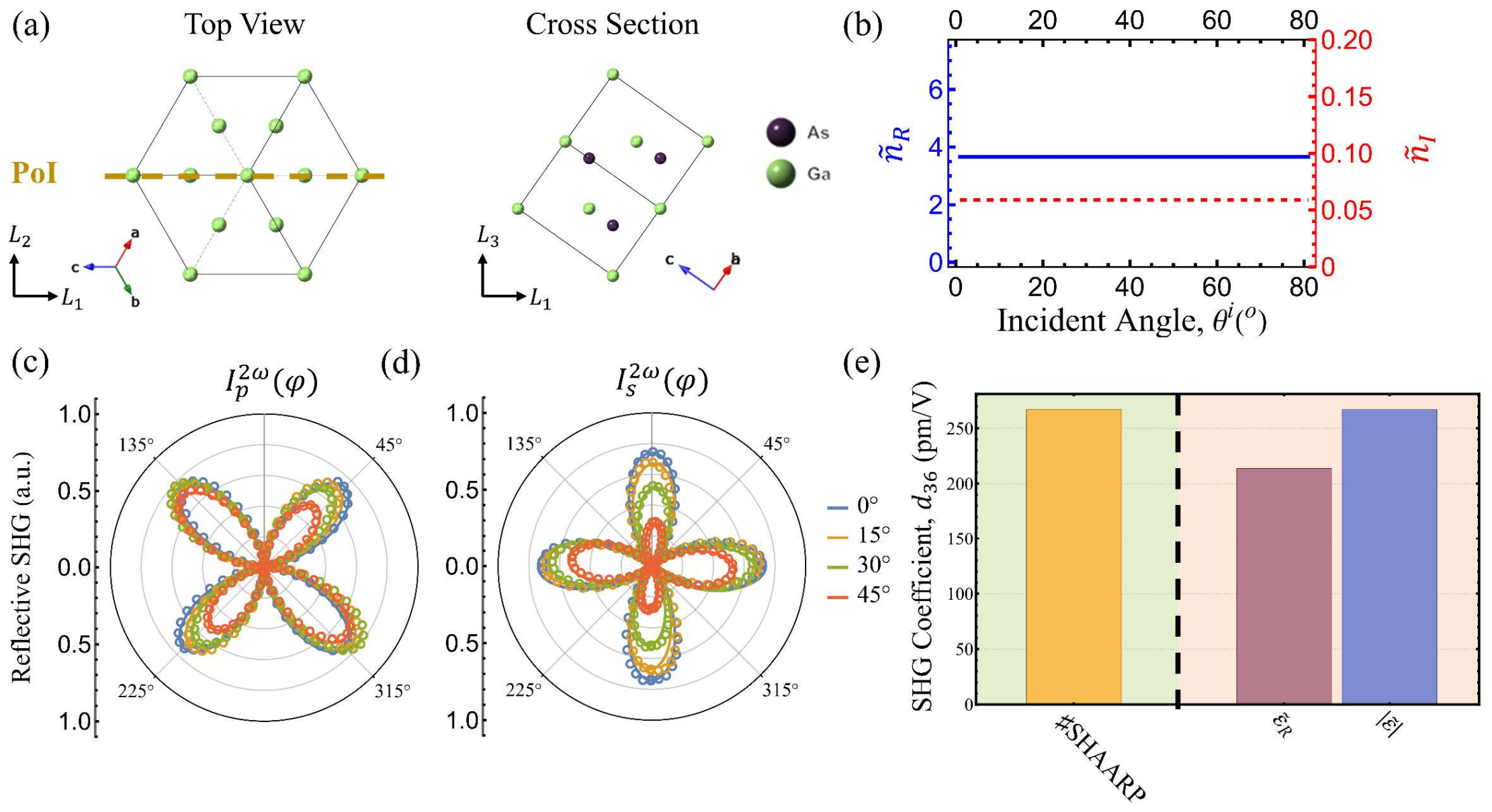

Figure 4. Anisotropic linear and nonlinear optical response of GaAs (111). **a** Crystal structure and experimental orientation. The $(L_1, L_2, L_3)$ and $(a, b, c)$ are the lab coordinate system and crystallographic coordinate system respectively. The dashed dark yellow line represents the projection of the plane of incidence to the $L_1 - L_2$ plane. **b** The effective complex refractive indices at $\omega$ frequency as a function of incident angle ($\theta^i$), subscript $R$ and $I$ represent real and imaginary respectively. **c-d** $p$- and $s$- polarized second harmonic response as a function of azimuthal angle ($\varphi$) at various incident angles ($\theta^i$). The open circles are experimental results, and the solid lines are theoretical fits. **e** Comparison of the SHG coefficients with no approximations (labeled as ♯SHAARP, pale green background) as against those extracted with various approximations (pale orange background). $\tilde{\boldsymbol{\varepsilon}}_R$, and $|\tilde{\boldsymbol{\varepsilon}}|$ represent the real component of $\tilde{\boldsymbol{\varepsilon}}$, and the magnitude of the complex $\tilde{\boldsymbol{\varepsilon}}$, respectively.

## Uniaxial $LiNbO_3$ $(11\bar{2}0)$

$LiNbO_3$ is a uniaxial crystal (point group $3m$) exhibiting ferroelectricity[38–40], piezoelectricity[41–43], and excellent nonlinear optical and electro-optic properties[44–46]. The nonlinear optical properties of $LiNbO_3$ have been well characterized using the Maker fringes technique based on the transmission geometry.[9,47] The bandgap of $LiNbO_3$ is reported to be ~3.78 eV[48,49], leading to no absorption at the fundamental pump energy of 1.55 eV and at the resulting SHG energy used in this study. Here we evaluate the absolute value of the SHG coefficient of $LiNbO_3$ by performing polarimetry experiments in the reflection geometry and using ♯SHAARP to fit the measured data. To exclude the contribution from the bottom surface of the $LiNbO_3$, a wedged crystal is used to select the contribution from only the top surface of the crystal.

The CCS can be expressed in hexagonal notation or trigonal notation, based on the choice of four or three basis vectors, respectively.[50] X-cut $LiNbO_3$ is used in this study, which has the surface plane $(11\bar{2}0)$ and surface normal along $[11\bar{2}0]$. The crystal is oriented in our measurement such that $[0001] \parallel L_1$ and $[11\bar{2}0] \parallel L_3$ as illustrated in **Fig.5a**. Since the optical axis is within the PoI, the extraordinary wave (transverse magnetic wave or the *p*-wave) experiences an effective refractive index as a function of the incidence angle ($\theta^i$) while the ordinary wave (transverse electric wave or the *s*-wave) experiences a constant refractive index, as shown in **Fig. 5b**. At $2\omega$ frequency, two homogeneous and three inhomogeneous waves propagate inside the crystal with distinct phase velocities. **Figure. 5c-d** show the *p*- and *s*- polarized second harmonic intensities, respectively, where open circles are experimental results and the solid curves are theoretical fitting from ♯SHAARP, respectively.

With the full consideration of anisotropic linear and nonlinear susceptibility and five nonlinear waves mixing, the fitting yields $\frac{d_{33}}{d_{31}} = 5.21 \pm 0.13$, which agrees well with the reported ratio ($\sim 5.3$) measured using the Maker fringes[47]. Further referencing against a $\alpha$-quartz (0.3 $d_{11}$ pm/V)[12,51,52] yields $d_{33}(\mathrm{LiNbO_3}) = 28.5 \pm 0.2$ pm/V, which shows excellent agreement compared with the reported value of $26.2 \pm 2.8$ pm/V[12]. To examine the impact of various fitting assumptions commonly made in the literature, we refit the same experimental data using the analytical expressions generated by ♯SHAARP and forcing certain assumptions. The assumptions of isotropic symmetry and Kleinman's symmetry were used to compare against the actual case as presented in **Fig. 5e**. Isotropic symmetry assumes isotropic dielectric permittivity and the absence of birefringence in the analysis. Here, we denote Iso($\varepsilon_{ii}$) as the isotropic assumption of dielectric tensors, where $\varepsilon_{ii}$ represents the dielectric components in the PCS. Kleinman symmetry[53–55] (abbreviated as KS) assumes that all three subscript indices of the SHG tensor (representing polarizations of the three waves) are fully permutable. This assumption has been widely used in determining nonlinear coefficients to reduce the number of independent variables[56], which leads to $d_{15} = d_{31}$ in $LiNbO_3$. With the assumption of isotropic dielectric permittivity, the ratios of coefficients exhibit an error within 10%. However, Kleinman symmetry can introduce 20% error even for the nonresonant fundamental and SHG frequencies used in this study. Similar effects have been observed in other structures.[57,58]

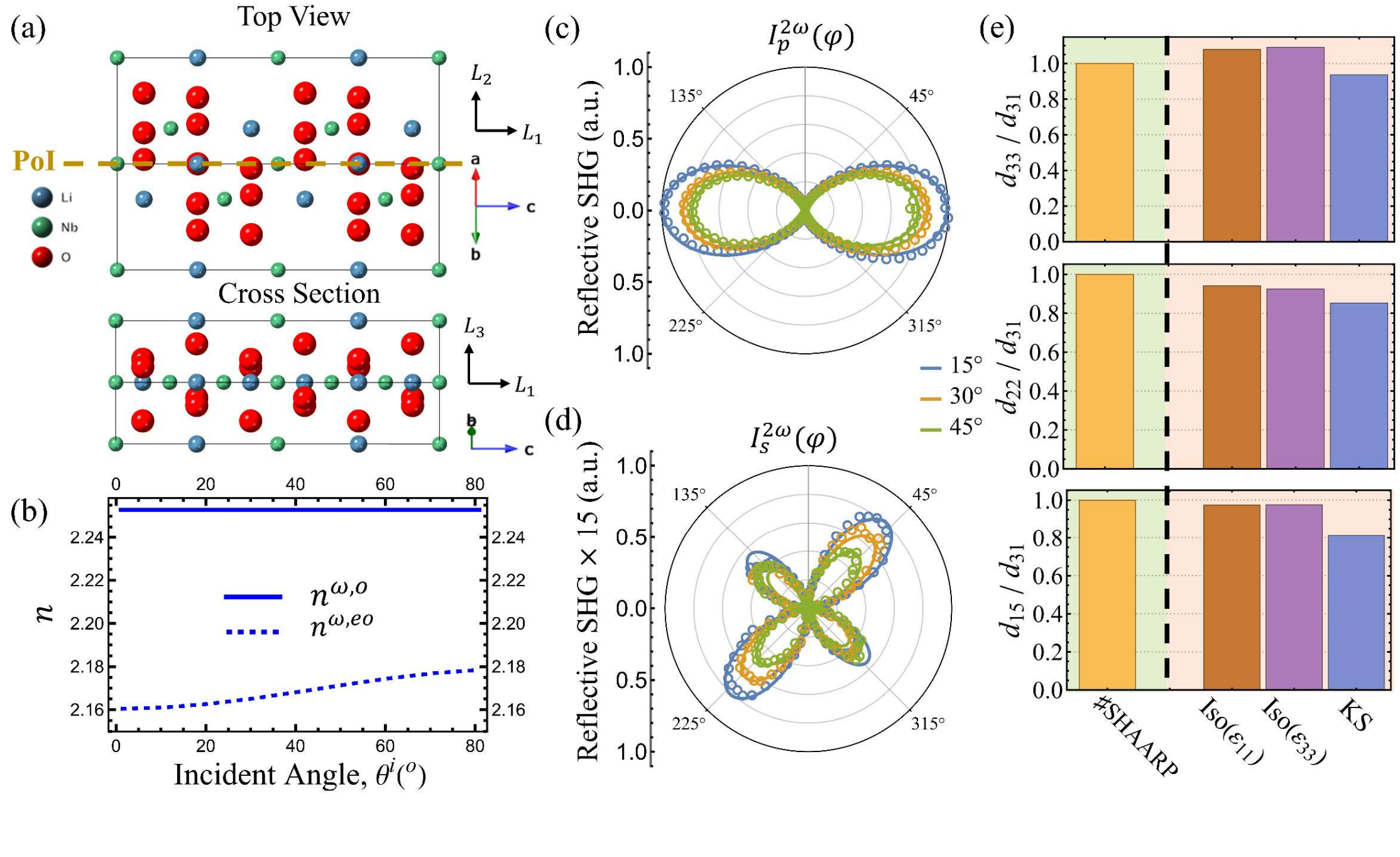

Figure 5. Linear and nonlinear optical response of $LiNbO_3$ $(11\bar{2}0)$. **a** Crystal structure and experimental orientation. **b** The real part of the effective refractive indices at $\omega$ frequency as a function of incident angle ($\theta^i$). **c-d** $p$- and $s$-polarized second harmonic response as a function of azimuthal angle ($\varphi$) at various incident angles ($\theta^i$). **e** Comparison of ratios of SHG coefficients between ♯SHAARP (light green background) and other forced approximations (light red background). Iso and KS respectively represent isotropic symmetry and Kleinman symmetry approximations, respectively. $Iso(\varepsilon_{ii})$ indicates that the indicated component $\varepsilon_{ii}$ was assumed in all directions under the isotropic approximation, where $i$ is the direction index.

## Biaxial KTP (100)

KTP is one of the most studied and widely applied NLO materials due to its large SHG coefficients and phase-matching properties.[59–62] KTP crystallizes in an orthorhombic structure with the point group *mm2* where the 2-fold axis is along the crystallographic *c*-axis.[63] The orthorhombic symmetry of KTP indicates that it belongs to the biaxial optical class characterized by three different refractive indices. The presence of two optical axes in such crystals often results in increased complexity in the analysis as compared with the uniaxial or isotropic classes.[23] KTP (100) is used for this study which is oriented so that $[001] \parallel L_2$ and $[100] \parallel L_3$ as shown in **Fig. 6a**. In this case, both the optical axes deviate from the PoI. The propagation and effective refractive indices of the extraordinary and ordinary waves can be obtained by solving the generalized eigenvalue problem as shown in **Fig. 6b**. Since the optical axes lie in the $L_2 - L_3$ plane, the two refractive waves are respectively polarized, one extraordinary and one ordinary, which can be decomposed as transverse electric (*s*-polarized) and transverse magnetic (*p*-polarized) waves.[64] The anisotropic SH signals reflected from the sample surface are collected at three incident angles ($\theta^i = 20°, 30°$, and $40°$). In total, we have sufficient sets of experimental data to uniquely determine all independent variables, by fitting the analytical expression obtained from ♯SHAARP. The fitting results are shown in **Figs. 6c-d**.

By using $LiNbO_3$ as the reference, the absolute $d_{33}$ is found to be $17.0 \pm 0.2$ pm/V which agrees well with values reported in the literature ($16.4 \pm 0.7$ pm/V)[12]. The extracted full tensor components (See **Table. S1**) are in excellent agreement with the previous study utilizing the Maker-fringe technique.[12] In **Fig. 6e** we compared the fitted ratios of nonlinear coefficients based on the full analysis with no approximations against the analysis performed by forcing various assumptions such as the isotropic dielectric permittivity and the Kleinman symmetry assumptions, which have been described above in the case of $LiNbO_3$. It is found that the errors in the acquired SHG coefficients introduced by these assumptions can reach up to 50% if dielectric tensors with higher symmetry are used.

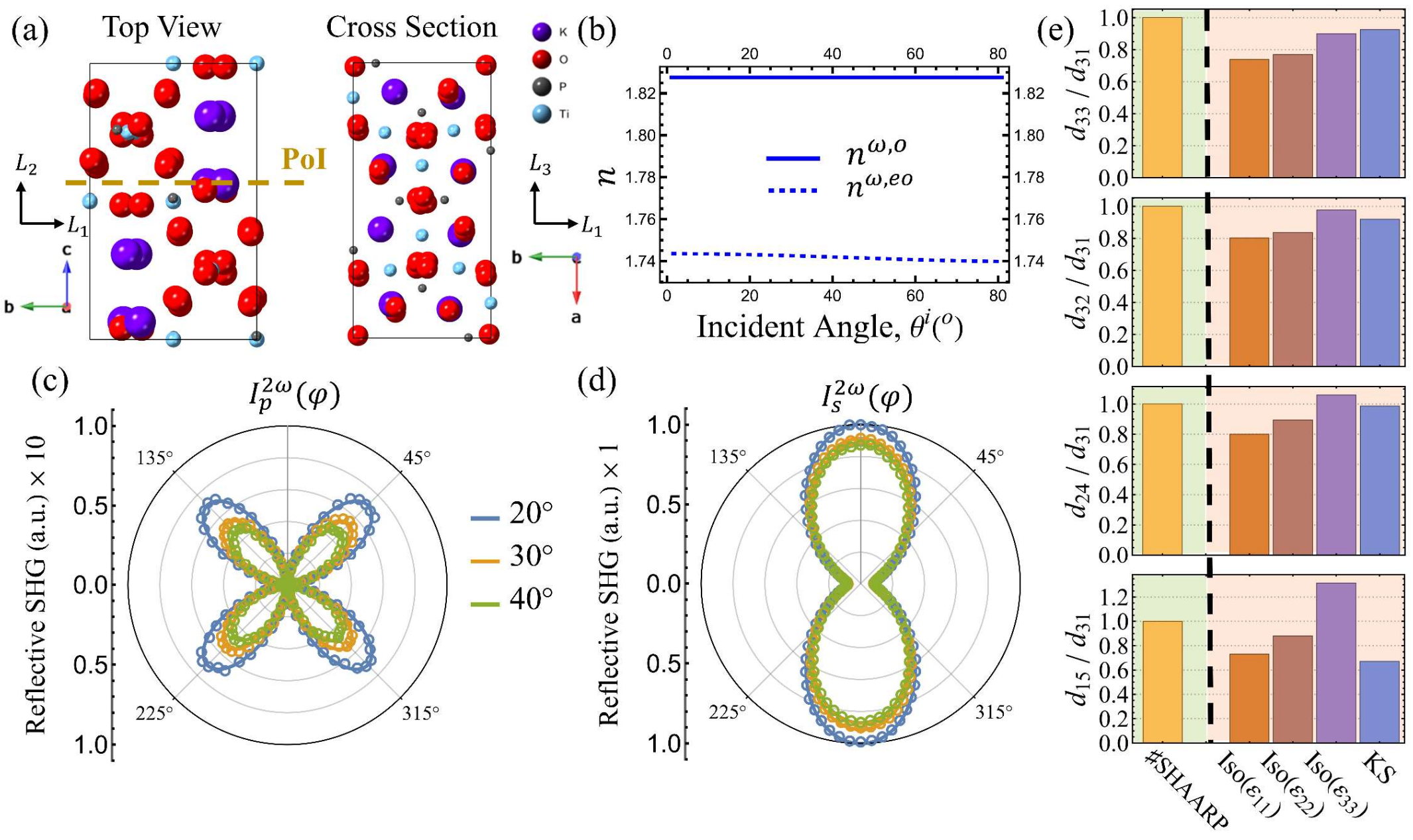

Figure 6. Linear and nonlinear optical response of KTP (100). **a** Crystal structure and experimental orientation. **b** The effective complex refractive indices at $\omega$ frequency as a function of incident angle ($\theta^i$). **c-d** *p*- and *s*- polarized second harmonic response as a function of azimuthal angle ($\varphi$) at various incident angles ($\theta^i$). **e** Comparison of the correct ratios of nonlinear coefficients (pale green background) as against other forced approximations (pale orange background). Iso, and KS respectively represent isotropic symmetry and Kleinman symmetry approximations,

respectively. $Iso(\varepsilon_{ii})$ indicates that the indicated component $\varepsilon_{ii}$ was assumed in all directions under the isotropic approximation, where $i$ is the direction index.

The excellent agreement between our results on KTP and those reported in the literature not only demonstrates the capability of analyzing biaxial crystals using the ♯SHAARP, but also that single surface reflection can be an effective approach to characterizing the SHG coefficients, complementing the widely used methods based on Maker fringes and the Bloembergen-Pershan relations.[9,10,14] All the extracted absolute SHG coefficients and ratios are summarized in **Table 1**.

## Weyl Semimetal, TaAs (112)

TaAs is one of the first experimentally identified Weyl semimetals which host intriguing transport properties.[65–68] Giant second harmonic response in TaAs was reported due to its intrinsic broken inversion symmetry and strong resonances.[21,69] Using ♯SHAARP, we measure and analyze the SHG coefficients of TaAs crystals and compare our results with those reported in the literature. In this example, we also highlight two unique features of ♯SHAARP, namely, the ability to incorporate absorption and arbitrary crystal surface orientation. As indicated in **Fig. 7a**, the surface plane is (112), and $[1\bar{1}0]$ is set parallel to the $L_2$. The (112)-oriented plane results in non-diagonalized dielectric tensor $\tilde{\boldsymbol{\varepsilon}}_r$ in the lab coordinate system and, consequently, the noncollinearity between $\boldsymbol{E}$ and $\boldsymbol{D}$. **Figure 7b** demonstrates the complex refractive index as a function of the incidence angle for both the ordinary and the extraordinary waves. The large extinction coefficient suggests large absorption at the pump wavelength. The anisotropic SHG intensities polarized along the $L_1$ and $L_2$ directions as a function of the polarization angle, $\varphi$, are measured, and then fitted with the analytical formula obtained using ♯SHAARP.

As seen in **Fig. 7c**, the intensity difference between $I_{L_1}^{2\omega}$ and $I_{L_2}^{2\omega}$ indicates a strong anisotropy of the second-order nonlinear optical properties of TaAs probed with fundamental energy at 1.55 eV. Due to its metallic nature, TaAs exhibits multiple resonances near $\omega$ and $2\omega$ frequencies leading to a resonance-enhanced SHG response.[69] By calibrating with the SH intensity of a well-studied nonlinear optical crystal ($LiNbO_3$ in our case), the absolute coefficients of TaAs can be uniquely determined. The saturation threshold of pump power was found to be around $50\mu W$(equivalent to a peak fluence of ~14 mJ/cm$^2$). The second harmonic response was then collected in the non-saturation regime where quadratic dependence between pump power and SH intensity remains (see **Fig. S1**). Using the equations generated by ♯SHAARP, the full absolute second harmonic tensor components can be evaluated, as shown in **Fig. 7d** (labeled as SHAARP).

With the full consideration of anisotropy, absorption, and boundary conditions at the interface, our analysis yields values for $d_{33}$ , $d_{31}$ and $d_{15}$ to be $827 \pm 39$, $12 \pm 15$, and $113 \pm 20$ pm/V, respectively. The strong anisotropy between $d_{33}$ and $d_{31}$ produces a large error bar for $d_{31}$. **Figure 7d** also demonstrates the variations of these coefficients created by forcing assumptions as shown in the pale orange-colored region. Here, Uni and Iso demonstrate the influence of birefringence on SHG response by constraining the dielectric tensors at both frequencies to be uniaxial and isotropic, respectively. To further explore the impact of optical resonances at the two frequencies on the SHG behavior, we adopted $\tilde{\boldsymbol{\varepsilon}}_R$ as the constraint using only real components of dielectric permittivities and $|\tilde{\boldsymbol{\varepsilon}}|$ as the magnitude of dielectric permittivities. The non-absorbing assumption when analyzing materials with strong resonance may significantly underestimate the intrinsic SHG coefficients as presented in Uni($\tilde{\boldsymbol{\varepsilon}}_R$) case. On the other hand, resonances due to interband transitions in TaAs[69] result in enhanced nonlinear optical susceptibilities, which make TaAs unsuitable for nonlinear optical frequency conversion applications due to its substantial absorption. Since utilizing photon energies below the bandgap is required for nonlinear optical frequency conversion applications, any direct comparison of TaAs or similar absorbing materials to transparent $LiNbO_3$ is not meaningful[70]. The amplitude of the overall dielectric permittivity (presented as Uni($|\tilde{\varepsilon}|$)) provides a closer estimation of the intrinsic properties as compared to the non-absorbing case. The study of Iso($\tilde{\varepsilon}_{ii}$) cases provides a comprehensive picture of the influence of birefringence in TaAs, where $\tilde{\varepsilon}_{ii}$ stands for the dielectric components used in the study. Iso($\tilde{\varepsilon}_{ii}$) refers to the cases that directly apply Bloembergen and Pershan formulae[8] to the studied material that has a symmetry lower than the cubic.[21] Due to the strong birefringence in TaAs, the isotropic dielectric permittivity assumption using $\tilde{\varepsilon}_{11}$ as the isotropic permittivity value (Iso($\tilde{\varepsilon}_{11}$)) leads to more than 20% underestimation of SHG coefficients, indicating the importance of a comprehensive and accurate model. Similarly, imposing Kleinman's symmetry (presented as KS) results in large variations in $d_{31}$ and $d_{15}$, and cannot be used as a simplification in TaAs. The surface normal of TaAs (112) is $[\frac{1}{a}\ \frac{1}{a}\ \frac{2}{c}]$ in the real space, where $a$ and $c$ are the lattice parameters[71]. The "Misoriented" case[21,22] in **Figure 7d** represents analysis using out-of-plane direction as $\left[\frac{1}{a}\ \frac{1}{a}\ \frac{2}{a}\right] = [112]$, assuming $a = c$. Due to the large $c/a$ ratio, the misorientation leads to a dramatic change in the analysis (see **Fig. S2**), and thus a large variation of the calculated $d_{33}$ in the analysis. Notably, the $d_{33}$ obtained without making these assumptions is found to be five times smaller as compared to the previously reported results.[21,22] It is found that this discrepancy most likely originates from a misorientation of the crystal used in a previously published model (detailed discussion in the Supplementary Note and **Fig. S2** where we reproduce the derivation[21,22] and identify the deviation from ♯SHAARP). Some differences due to the physical crystals are also possible, but the crystal used in this study shows that other properties of the crystals are comparable.[69] Our results reveal that nothing more than a misorientation in modeling can lead to a huge overestimation of $d_{33}$ further emphasizing the importance of software tools such as ♯SHAARP that can eliminate such inadvertent human errors.

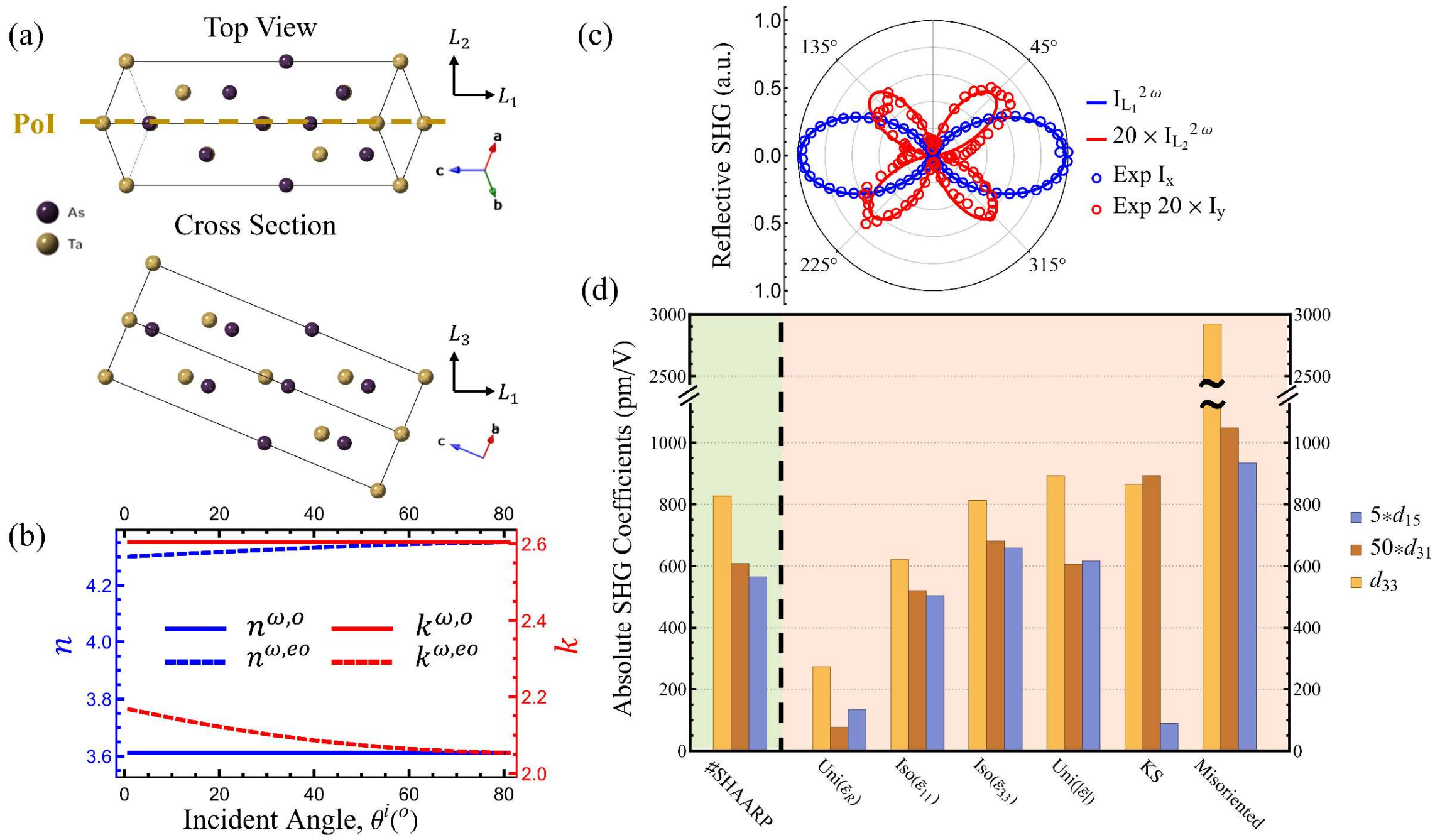

Figure 7. Linear and nonlinear optical response of TaAs (112). **a** Crystal structure and experimental orientation. **b** The effective complex refractive indices at $\omega$ frequency as a function of incident angle ($\theta^i$). **c** *p*- and *s*- polarized second harmonic response as a function of azimuthal angle ($\varphi$) at various incident angles ($\theta^i$). **e** Comparison of the correct ratios of nonlinear coefficients (pale green background) as against other forced approximations (pale orange background). Uni, Iso, and KS respectively represent uniaxial, isotropic symmetry, and Kleinman's symmetry approximations. $Iso(\varepsilon_{ii})$ indicates that the indicated component $\varepsilon_{ii}$ was assumed in all directions under the isotropic approximation, where $i$ is the direction index. Uni($|\tilde{\boldsymbol{\varepsilon}}|$) and Uni($\tilde{\boldsymbol{\varepsilon}}_R$) respectively mean using the magnitudes of complex relative dielectric permittivity and the real part of the dielectric permittivity as tensor values while retaining the uniaxial dielectric permittivity tensor symmetry. Misoriented case represents analysis using the plane normal as [112] which is discussed in detail in the supplementary.

# Discussion

**Table. 1** summarizes absolute SHG coefficients and SHG ratios from this study and literature. We have benchmarked our analysis using three well-known nonlinear optical materials (GaAs, $LiNbO_3$, and KTP), covering all three optical classes, both transparent and absorbing systems, and both optical axes within and away from PoI. Excellent agreement has been achieved between this work and literature for those three classical materials. We have examined how different assumptions made to simplify the analytical formulae of SHG intensities can influence the fitting results and thus the accuracy of the absolute values of SHG coefficients for the four different nonlinear optical crystals shown above. Overall in these examples, we found that including absorption and accurate crystal orientation, among other considerations, play significant roles in determining the nonlinear optical coefficients. Simply using the magnitude of the complex dielectric permittivity may result in an error of less than 10%, but neglecting extinction coefficients entirely may generate significant errors in the analysis (up to 90% for $d_{31}$ in TaAs). Simplifying the analysis by assuming higher symmetry of dielectric permittivity could introduce errors up to 20% or more, depending on the optical birefringence. However, assuming Kleinman's symmetry results in much larger errors as compared to assuming higher symmetry for the cases presented in this study.

Table 1. Comparison of ratios of SHG coefficients from ♯SHAARP and literature. Absolute values are in the unit of pm/V.

| Materials | SHG Coefficients | This work | Ref |
|---|---|---|---|
| GaAs | $\|d_{36}\|$ | $267 \pm 20$ | $310 \pm 50$[35] |
| $LiNbO_3$ | $\|d_{33}\|$ | $28.5 \pm 0.2$ | $26.2 \pm 2.8$[48,60] |
| | $d_{33}/d_{31}$ | $5.2 \pm 0.1$ | $5.35 \pm 0.44$[48,60] |
| | $d_{22}/d_{31}$ | $-0.23 \pm 0.02$ | $-0.49 \pm 0.11$[48,60] |
| KTP | $\|d_{33}\|$ | $17.0 \pm 0.2$ | $16.4 \pm 0.7$[48,60] |
| | $d_{33}/d_{31}$ | $6.0 \pm 0.4$ | $6.7 \pm 0.6$[48,60] |
| | $d_{32}/d_{31}$ | $1.6 \pm 0.1$ | $1.7 \pm 0.2$[48,60] |
| | $d_{24}/d_{31}$ | $1.5 \pm 0.1$ | $1.7 \pm 0.2$[48,60] |
| | $d_{15}/d_{31}$ | $1.3 \pm 0.3$ | $0.8 \pm 0.1$[48,60] |
| TaAs | $\|d_{33}\|$ | $\pm 827 \pm 39$ | $3600 \pm 550$[21] |

| | |
|---|---|
| $\lvert d_{31} \rvert$ | $\pm 12 \pm 15$ |
| $\lvert d_{15} \rvert$ | $\pm 113 \pm 20$ |

♯SHAARP demonstrates accurate and reliable analysis comparable with the Maker fringes, which have been used for six decades for the analysis of second harmonic response from transparent crystals.[9,10,12,14] Maker fringes method is powerful for characterizing nonlinear susceptibilities especially in analyzing transparent, thin crystals with high symmetry orientations along the LCS. However, the walk-off angle, the requirement of transmission geometry, simplified orientation, and high symmetry have limited the broader applications of this method. Moreover, the Maker fringes method mainly focuses on *p*- and *s*- polarized incident and SHG waves limiting the tunability of the detection schemes. On the other hand, we have demonstrated that the single-surface reflection method is as robust as the Maker fringes technique. This method provides more flexibility in materials selection regardless of their orientation, thickness, absorption, and symmetry. Polarimetry obtained through this method maps out the complete dependence of both input and output polarization, providing sufficient information on the nonlinear susceptibility tensor. Furthermore, various combinations of polarization settings provided in ♯SHAARP, including linear, elliptical, and circular polarized light, could promote more experimental designs such as for chiral structures.

## Summary

We developed an open-source package, ♯SHAARP, for simulating SHG responses and extracting intrinsic SHG coefficients of nonlinear optical crystals based on a single-interface reflection geometry. The package is generally applicable to analyzing the reflective SHG of materials with arbitrary crystal symmetry, surface orientation, and absorption. To benchmark the results obtained by ♯SHAARP, we performed polarimetry experiments on representative nonlinear optical crystals, including GaAs, $LiNbO_3$, and KTP, to measure the reflective SHG intensities and fit the measurements with the analytical formula obtained ♯SHAARP. We extracted the absolute SHG coefficients of the three materials which show excellent quantitative agreement with the previous works based on the transmission geometry using the Maker-fringe technique. We further applied ♯SHAARP to evaluate the SHG coefficients of topological Weyl semimetal TaAs. We found that the resonant SHG coefficient $d_{33}$ is nearly five times smaller than that reported in previous literature[21,22]. Possible reasons for this deviation were discussed.

Looking forward, we believe that the open-source software, ♯SHAARP, will benefit studies of nonlinear optical materials in numerical modeling the polarimetry of known materials and extracting the absolute SHG coefficients of new materials. Moreover, the experimental scheme based on the single-interface reflection geometry will provide an alternative, more flexible way to the Maker fringe method in evaluating the SHG coefficients of nonlinear optical crystals, especially when the transmission experiment is challenging, e.g., for absorbing crystals. Meanwhile, we are building new functionalities into ♯SHAARP to enable the numerical and analytical modeling of the SHG based on transmission geometry through slabs and the Maker fringes. A version of ♯SHAARP for multiple-interface or multilayer system is under development.

## Methods

Sample Preparation: GaAs and $LiNbO_3$ single crystals were obtained from MTI Corporation. Note that X-cut based on MTI definition is different compared to convention[25,72] and we have specified the $LiNbO_3$ $(11\bar{2}0)$ for clarity. The KTP crystal was obtained from CASTECH Inc. TaAs (112) was grown by chemical vapor transport at ~1000°C for four weeks. Details can be found in the previous work[69]. Wedged X-cut $LiNbO_3$ was prepared using 10×10×1 mm crystal with a wedged angle of around 5 degrees. The crystal was optically polished with $0.05\mu m$ alumina suspension.

Second-harmonic generation: Second-harmonic polarimetry was performed using a Ti: Sapphire femtosecond laser system with the central wavelength at 800 nm (1 kHz, 100 fs). The incident linear polarized light was rotated through a zero-order half-wave plate and focused on the sample surface. A collecting lens was placed at the reflection geometry and the SH signals were filtered by an analyzer and a bandpass filter before entering the photomultiplier tube (PMT). The SH signals were then processed by the lock-in amplifier to filter out noise. The SHG fittings were then conducted using the expression generated by the ♯SHAARP. All the SHG coefficients from the literature are recalibrated using Miller's rule before the comparison.[73]

## Acknowledgment

This development of the software was supported as part of the Computational Materials Sciences Program funded by the U.S. Department of Energy, Office of Science, Basic Energy Sciences, under Award No. DE-SC0020145. R.Z., B.W., L.Q.C and V.G. were supported by U.S. Department of Energy, Office of Science, Basic Energy Sciences, under Award No. DE-SC0020145. J.H. acknowledges Air Force Office of Scientific Research Grant number FA9550-18-S-0003. L.W. was supported by NSF Research Experiences for Undergraduates (REU), DMR 185-1987.

## Competing Interest

The authors declare no competing interests.

## Data Availability

The data that support the findings of this study are available from the corresponding author upon reasonable request.

The ♯SHAARP.*si* is available through GitHub (https://github.com/Rui-Zu/SHAARP) and the documentation of the ♯SHAARP.*si* can be accessed through ReadtheDocs (https://shaarp.readthedocs.io/en/latest/).

## Reference

1. Picqué, N. & Hänsch, T. W. Mid-IR spectroscopic sensing. *Optics and Photononics News* 26–33 (2019).
2. Kaushal, H. & Kaddoum, G. Optical Communication in Space: Challenges and Mitigation Techniques. *IEEE Communications Surveys & Tutorials* **19**, 57–96 (2017).
3. Mansell, G. L. *et al.* Observation of Squeezed Light in the <math display="inline"> <mrow> <mn>2</mn> <mtext> </mtext> <mtext> </mtext> <mi>μ</mi> <mi mathvariant="normal">m</mi> </mrow> </math> Region. *Physical Review Letters* **120**, 203603 (2018).
4. Fülöp, J. A., Tzortzakis, S. & Kampfrath, T. Laser-Driven Strong-Field Terahertz Sources. *Advanced Optical Materials* **8**, 1–25 (2020).
5. Blanchard, F., Doi, A., Tanaka, T. & Tanaka, K. Real-Time, Subwavelength Terahertz Imaging. *Annual Review of Materials Research* **43**, 237–259 (2013).
6. Kimble, H. J. The quantum internet. *Nature* **453**, 1023–1030 (2008).
7. Newnham, R. E. *Properties of Materials: Anisotropy, Symmetry, Structure*. (OUP Oxford, 2005).
8. Bloembergen, N. & Pershan, P. S. Light Waves at the Boundary of Nonlinear Media. *Phys. Rev.* **128**, 606–622 (1962).
9. Maker, P. D., Terhune, R. W., Nisenoff, M. & Savage, C. M. Effects of Dispersion and Focusing on the Production of Optical Harmonics. *Phys. Rev. Lett.* **8**, 21–22 (1962).
10. Herman, W. N. & Hayden, L. M. Maker fringes revisited: second-harmonic generation from birefringent or absorbing materials. *J. Opt. Soc. Am. B, JOSAB* **12**, 416–427 (1995).

11. Santos, D. F., Guerreiro, A. & Baptista, J. M. Numerical investigation of a refractive index SPR D-type optical fiber sensor using COMSOL multiphysics. *Photonic Sens* **3**, 61–66 (2013).

12. Shoji, I., Kondo, T., Kitamoto, A., Shirane, M. & Ito, R. Absolute scale of second-order nonlinear-optical coefficients. *J. Opt. Soc. Am. B, JOSAB* **14**, 2268–2294 (1997).

13. Sipe, J. E., Moss, D. J. & van Driel, H. M. Phenomenological theory of optical second- and third-harmonic generation from cubic centrosymmetric crystals. *Phys. Rev. B* **35**, 1129–1141 (1987).

14. Jerphagnon, J. & Kurtz, S. K. Maker Fringes: A Detailed Comparison of Theory and Experiment for Isotropic and Uniaxial Crystals. *Journal of Applied Physics* **41**, 1667–1681 (1970).

15. Qian, Q. *et al.* Chirality-Dependent Second Harmonic Generation of MoS2 Nanoscroll with Enhanced Efficiency. *ACS Nano* **14**, 13333–13342 (2020).

16. Lei, S. *et al.* Observation of Quasi-Two-Dimensional Polar Domains and Ferroelastic Switching in a Metal, $Ca_3Ru_2O_7$. *Nano Letters* **18**, 3088–3095 (2018).

17. Dou, S. X., Jiang, M. H., Shao, Z. S. & Tao, X. T. Maker fringes in biaxial crystals and the nonlinear optical coefficients of thiosemicarbazide cadmium chloride monohydrate. *Appl. Phys. Lett.* **54**, 1101–1103 (1989).

18. Nye, J. F. *Physical Properties of Crystals: Their Representation by Tensors and Matrices*. (Oxford University Press, 1985).

19. Kruk, S. *et al.* Enhanced Magnetic Second-Harmonic Generation from Resonant Metasurfaces. *ACS Photonics* **2**, 1007–1012 (2015).

20. Oskooi, A. F. *et al.* Meep: A flexible free-software package for electromagnetic simulations by the FDTD method. *Computer Physics Communications* **181**, 687–702 (2010).

21. Wu, L. *et al.* Giant anisotropic nonlinear optical response in transition metal monopnictide Weyl semimetals. *Nature Physics* **13**, 350–355 (2017).

22. Patankar, S. *et al.* Resonance-enhanced optical nonlinearity in the Weyl semimetal TaAs. *Phys. Rev. B* **98**, 165113 (2018).

23. Yariv, A., Yariv, P. of E. E. A., Yeh, P. & Yeh, P. *Optical Waves in Crystals: Propagation and Control of Laser Radiation*. (Wiley, 1984).

24. Newnham, R. E. *Properties of Materials: Anisotropy, Symmetry, Structure*. (OUP Oxford, 2005).

25. IEEE Standard on Piezoelectricity. *ANSI/IEEE Std 176-1987* 0_1- (1988) doi:10.1109/IEEESTD.1988.79638.

26. Trolier-McKinstry, S. & Newnham, R. E. *Materials Engineering: Bonding, Structure, and Structure-Property Relationships*. (Cambridge University Press, 2017).

27. Yariv, A. *Optical Waves in Crystals: Propagation and Controlof Laser Radiation*. (Wiley-Interscience, 2002).

28. Born, M. & Wolf, E. *Principles of Optics: Electromagnetic Theory of Propagation, Interference and Diffraction of Light*. (Cambridge University Press, 1999). doi:10.1017/CBO9781139644181.

29. Chang, C.-M. & Shieh, H.-P. D. Simple Formulas for Calculating Wave Propagation and Splitting in Anisotropic Media. *Jpn. J. Appl. Phys.* **40**, 6391–6395 (2001).

30. Masselink, W. T., Chang, Y.-C. & Morkoç, H. Acceptor spectra of Al x Ga 1 − x As-GaAs quantum wells in external fields: Electric, magnetic, and uniaxial stress. *Phys. Rev. B* **32**, 5190–5201 (1985).

31. Aspnes, D. E., Kelso, S. M., Logan, R. A. & Bhat, R. Optical properties of Al $_{x}$ Ga $_{1-x}$ As. *Journal of Applied Physics* **60**, 754–767 (1986).

32. Adachi, S. *Optical Constants of Crystalline and Amorphous Semiconductors*. (Springer US, 1999). doi:10.1007/978-1-4615-5247-5.

33. Blakemore, J. S. Semiconducting and other major properties of gallium arsenide. *Journal of Applied Physics* **53**, R123–R181 (1982).

34. Schunemann, P. G., Pomeranz, L. A., Young, Y. E., Mohnkern, L. & Vera, A. Recent advances in all-epitaxial growth and properties of orientation-patterned gallium arsenide (OP-GaAs). in *2009 Conference on Lasers and Electro-Optics and 2009 Conference on Quantum electronics and Laser Science Conference* 1–2 (2009). doi:10.1364/CLEO.2009.CWJ5.

35. Bergfeld, S. & Daum, W. Second-Harmonic Generation in GaAs: Experiment versus Theoretical Predictions of χ x y z ( 2 ). *Phys. Rev. Lett.* **90**, 036801 (2003).

36. Shan, Y. *et al.* Stacking symmetry governed second harmonic generation in graphene trilayers. *Science Advances* **4**, eaat0074 (2018).

37. Clark, D. J. *et al.* Strong optical nonlinearity of CVD-grown ${\mathrm{MoS}}_{2}$ monolayer as probed by wavelength-dependent second-harmonic generation. *Phys. Rev. B* **90**, 121409 (2014).

38. Nassau, K., Levinstein, H. J. & Loiacono, G. M. Ferroelectric lithium niobate. 1. Growth, domain structure, dislocations and etching. *Journal of Physics and Chemistry of Solids* **27**, 983–988 (1966).

39. *Scanning probe microscopy: electrical and electromechanical phenomena at the nanoscale*. (Springer, 2007).

40. *Physics of ferroelectrics: a modern perspective*. (Springer, 2007).

41. Weis, R. S. & Gaylord, T. K. Lithium niobate: Summary of physical properties and crystal structure. *Appl. Phys. A* **37**, 191–203 (1985).

42. Warner, A. W., Onoe, M. & Coquin, G. A. Determination of Elastic and Piezoelectric Constants for Crystals in Class (3 *m* ). *The Journal of the Acoustical Society of America* **42**, 1223–1231 (1967).

43. Cho, Y. & Yamanouchi, K. Nonlinear, elastic, piezoelectric, electrostrictive, and dielectric constants of lithium niobate. *Journal of Applied Physics* **61**, 875–887 (1987).

44. Miller, R. C., Nordland, W. A. & Bridenbaugh, P. M. Dependence of Second-Harmonic-Generation Coefficients of LiNbO3 on Melt Composition. *Journal of Applied Physics* **42**, 4145–4147 (1971).

45. Nikogosyan, D. N. *Nonlinear Optical Crystals: A Complete Survey*. (Springer-Verlag, 2005). doi:10.1007/b138685.

46. Sánchez-Dena, O. *et al.* Effect of size and composition on the second harmonic generation from lithium niobate powders at different excitation wavelengths. *Mater. Res. Express* **4**, 035022 (2017).

47. Shoji, I., Kondo, T., Kitamoto, A., Shirane, M. & Ito, R. Absolute scale of second-order nonlinear-optical coefficients. *J. Opt. Soc. Am. B, JOSAB* **14**, 2268–2294 (1997).

48. Nikogosyan, D. N. *Nonlinear Optical Crystals: A Complete Survey*. (Springer-Verlag, 2005). doi:10.1007/b138685.

49. Dhar, A. & Mansingh, A. Optical properties of reduced lithium niobate single crystals. *Journal of Applied Physics* **68**, 5804–5809 (1990).

50. Graef, M. D. & McHenry, M. E. *Structure of Materials: An Introduction to Crystallography, Diffraction and Symmetry*. (Cambridge University Press, 2012).

51. Hagimoto, K. & Mito, A. Determination of the second-order susceptibility of ammonium dihydrogen phosphate and α-quartz at 633 and 1064 nm. *Appl. Opt., AO* **34**, 8276–8282 (1995).

52. Roberts, D. A. Simplified characterization of uniaxial and biaxial nonlinear optical crystals: a plea for standardization of nomenclature and conventions. *IEEE Journal of Quantum Electronics* **28**, 2057–2074 (1992).

53. Kleinman, D. A. Nonlinear Dielectric Polarization in Optical Media. *Phys. Rev.* **126**, 1977–1979 (1962).

54. Dailey, C. A., Burke, B. J. & Simpson, G. J. The general failure of Kleinman symmetry in practical nonlinear optical applications. *Chemical Physics Letters* **390**, 8–13 (2004).

55. Miller, R. C. OPTICAL SECOND HARMONIC GENERATION IN PIEZOELECTRIC CRYSTALS. *Appl. Phys. Lett.* **5**, 17–19 (1964).

56. Okada, M. & Ieiri, S. Kleinman's symmetry relation in non-linear optical coefficient of LiIO3. *Physics Letters A* **34**, 63–64 (1971).

57. Zhao, H.-J., Zhang, Y.-F. & Chen, L. Strong Kleinman-Forbidden Second Harmonic Generation in Chiral Sulfide: $La_4InSbS_9$. *J. Am. Chem. Soc.* **134**, 1993–1995 (2012).

58. Chemla, D. S. & Jerphagnon, J. Optical Second-Harmonic Generation in Paratellurite and Kleinman's Symmetry Relations. *Appl. Phys. Lett.* **20**, 222–223 (1972).

59. Sturm, C., Zviagin, V. & Grundmann, M. Dielectric tensor, optical activity, and singular optic axes of KTP in the spectral range 0.5–8.4 eV. *Phys. Rev. Materials* **4**, 055203 (2020).

60. Shoji, I., Kondo, T., Kitamoto, A., Shirane, M. & Ito, R. Absolute scale of second-order nonlinear-optical coefficients. *J. Opt. Soc. Am. B, JOSAB* **14**, 2268–2294 (1997).

61. Fan, T. Y. *et al.* Second harmonic generation and accurate index of refraction measurements in flux-grown $KTiOPO_4$. *Appl. Opt., AO* **26**, 2390–2394 (1987).

62. Vanherzeele, H. & Bierlein, J. D. Magnitude of the nonlinear-optical coefficients of KTiOPO_4. *Opt. Lett.* **17**, 982 (1992).

63. Dahaoui, S., Hansen, N. K. & Menaert, B. NaTiOPO4 and KTiOPO4 at 110K. *Acta Cryst C* **53**, 1173–1176 (1997).

64. Okamoto, K. *Fundamentals of Optical Waveguides*. (Academic Press, 2005).

65. Yuan, X. *et al.* The discovery of dynamic chiral anomaly in a Weyl semimetal NbAs. *Nature Communications* **11**, 1259 (2020).

66. Leahy, I. A. *et al.* Nonsaturating large magnetoresistance in semimetals. *Proc Natl Acad Sci USA* **115**, 10570–10575 (2018).

67. Yang, L. X. *et al.* Weyl semimetal phase in the non-centrosymmetric compound TaAs. *Nature Physics* **11**, 728–732 (2015).

68. Huang, S.-M. *et al.* A Weyl Fermion semimetal with surface Fermi arcs in the transition metal monopnictide TaAs class. *Nature Communications* **6**, 7373 (2015).

69. Zu, R. *et al.* Comprehensive anisotropic linear optical properties of the Weyl semimetals TaAs and NbAs. *Phys. Rev. B* **103**, 165137 (2021).

70. Abdelwahab, I. *et al.* Giant second-harmonic generation in ferroelectric NbOI2. *Nat. Photon.* (2022) doi:10.1038/s41566-022-01021-y.

71. Furuseth, S. *et al.* On the Arsenides and Antimonides of Tantalum. *Acta Chem. Scand.* **19**, 95–106 (1965).

72. Sanna, S. & Schmidt, W. G. $LiNbO_3$ surfaces from a microscopic perspective. *J. Phys.: Condens. Matter* **29**, 413001 (2017).

73. Boyd, R. W. & Prato, D. *Nonlinear Optics*. (Academic Press, 2008).

Supporting Information

# ♯SHAARP: An Open-Source Package for Analytical and Numerical Modeling of Optical Second Harmonic Generation in Anisotropic Crystals

Rui Zu[1,a], Bo Wang[1,5,a], Jingyang He[1], Jian-Jun Wang[1], Lincoln Weber[1], Long-Qing Chen[1,3,4], Venkatraman Gopalan[1,2,3]

[a]These authors contribute equally to this work

[1]Department of Materials Science and Engineering, The Pennsylvania State University, University Park, Pennsylvania 16802, USA

[2] Department of Physics, Pennsylvania State University, University Park, Pennsylvania, 16802, USA

[3]Department of Engineering Science and Mechanics, The Pennsylvania State University, University Park, Pennsylvania 16802, USA

[4]Department of Mathematics, The Pennsylvania State University, University Park, Pennsylvania 16802, USA

[5]Materials Science Division, Lawrence Livermore National Laboratory, Livermore, CA 94550, USA

Venkatraman Gopalan (vxg8@psu.edu); Long-Qing Chen (lqc3@psu.edu); Rui Zu (ruizu0110@gmail.com); Bo Wang (bzw133.psu@gmail.com)

## 1. Power-dependent second harmonic generation of TaAs (112)

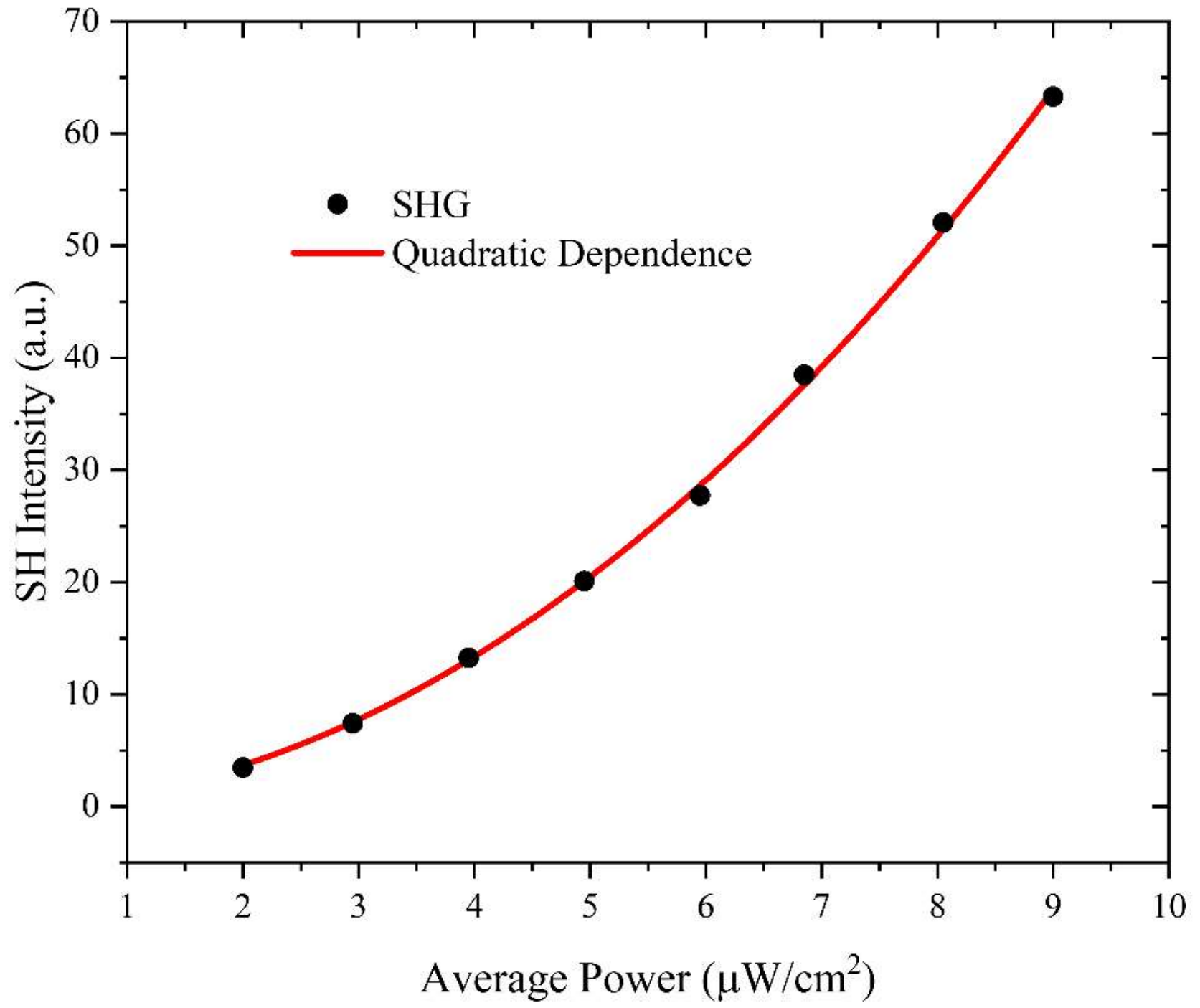

Fig S1. Power-dependent SHG response of TaAs (112). The dot and line correspond to data points and quadratic fitting.

## 2. SHG analysis from TaAs (112) surface

TaAs crystallizes in a tetragonal structure with point group $4mm$. Unlike the cubic structure, the length of the four-fold rotation axis of TaAs (crystallographic c axis) is distinct from its $a$ and $b$ axes. Therefore, the $[\frac{1}{a}\ \frac{1}{a}\ \frac{2}{a}]$=[112] direction and the $[\frac{1}{a}\ \frac{1}{a}\ \frac{2}{c}]$ direction (surface normal of TaAs (112)) are different. By converting (112) in the reciprocal space to real space, the crystal physics axes $(Z_1, Z_2, Z_3)$ in the lab coordinate system are found to be (-0.272,0.707,0.653), (-0.272,-0.707,0.653) and (0.923,0,0.385) as shown in **Fig. S2a**. On the other hand, if we mistakenly assume [1 1 -1], [1 -1 0], and [1 1 2] as the directions parallel to $L_1$, $L_2$ and $L_3$ respectively, ignoring the fact that $c$ is around 3 times larger than $a$ and $b$ axis, the crystal physics axes in the lab coordinate system are $\left(-\frac{1}{\sqrt{3}}, -\frac{1}{\sqrt{2}}, \frac{1}{\sqrt{6}}\right)$, $\left(-\frac{1}{\sqrt{3}}, \frac{1}{\sqrt{2}}, \frac{1}{\sqrt{6}}\right)$, and $\left(\frac{1}{\sqrt{3}}, 0, -\sqrt{\frac{2}{3}}\right)$ as shown in **Fig. S2b**. Based on the misorientation, the resulting (incorrect) SHG expressions are given below, which are the same as those derived in the previous literature.[1,2]

$$I_{L_1}^{2\omega} = \frac{1}{27}((4d_{15} + 2d_{31} + d_{33})\cos^2\varphi + 3d_{31}\sin^2\varphi)^2 \tag{S1}$$

$$I_{L_2}^{2\omega} = \frac{1}{3}{d_{15}}^2\sin^2 2\varphi \tag{S2}$$

$$I_{\parallel}^{2\omega} = \frac{1}{27}((4d_{15} + 2d_{31} + d_{33})\cos^3\varphi + 3(2d_{15} + d_{31})\sin^2\varphi\cos\varphi)^2 \tag{S3}$$

$$I_{\perp}^{2\omega} = \frac{1}{27}((-2d_{15} + 2d_{31} + d_{33})\cos^2\varphi\sin\varphi + 3d_{31}\sin^3\varphi)^2 \tag{S4}$$

The correct SHG expressions for the TaAs(112) surface are given below:

$$I_{L_1}^{2\omega} = \frac{c^2(d_{31}(2a^2+c^2)\sin^2(\varphi)+\cos^2(\varphi)(2a^2(2d_{15}+d_{31})+c^2d_{33}))^2}{(2a^2+c^2)^3} \tag{S5}$$

$$I_{L_2}^{2\omega} = \frac{c^2d_{15}^2\sin^2(2\varphi)}{2a^2+c^2} \tag{S6}$$

$$I_{\parallel}^{2\omega} = \frac{c^2\cos^2(\varphi)((2d_{15}+d_{31})(2a^2+c^2)\sin^2(\varphi)+\cos^2(\varphi)(2a^2(2d_{15}+d_{31})+c^2d_{33}))^2}{(2a^2+c^2)^3} \tag{S7}$$

$$I_{\perp}^{2\omega} = \frac{c^2\sin^2(\varphi)(d_{31}(2a^2+c^2)\sin^2(\varphi)+\cos^2(\varphi)(2a^2d_{31}+c^2(d_{33}-2d_{15})))^2}{(2a^2+c^2)^3} \tag{S8}$$

Note that derivations mentioned above do not involve boundary condition analysis. The derivations are simply derived from $I^{2\omega} \propto (\boldsymbol{P}^{NL})^2 = (\boldsymbol{d}^{SHG}\boldsymbol{E}^{\omega}\boldsymbol{E}^{\omega})^2$.

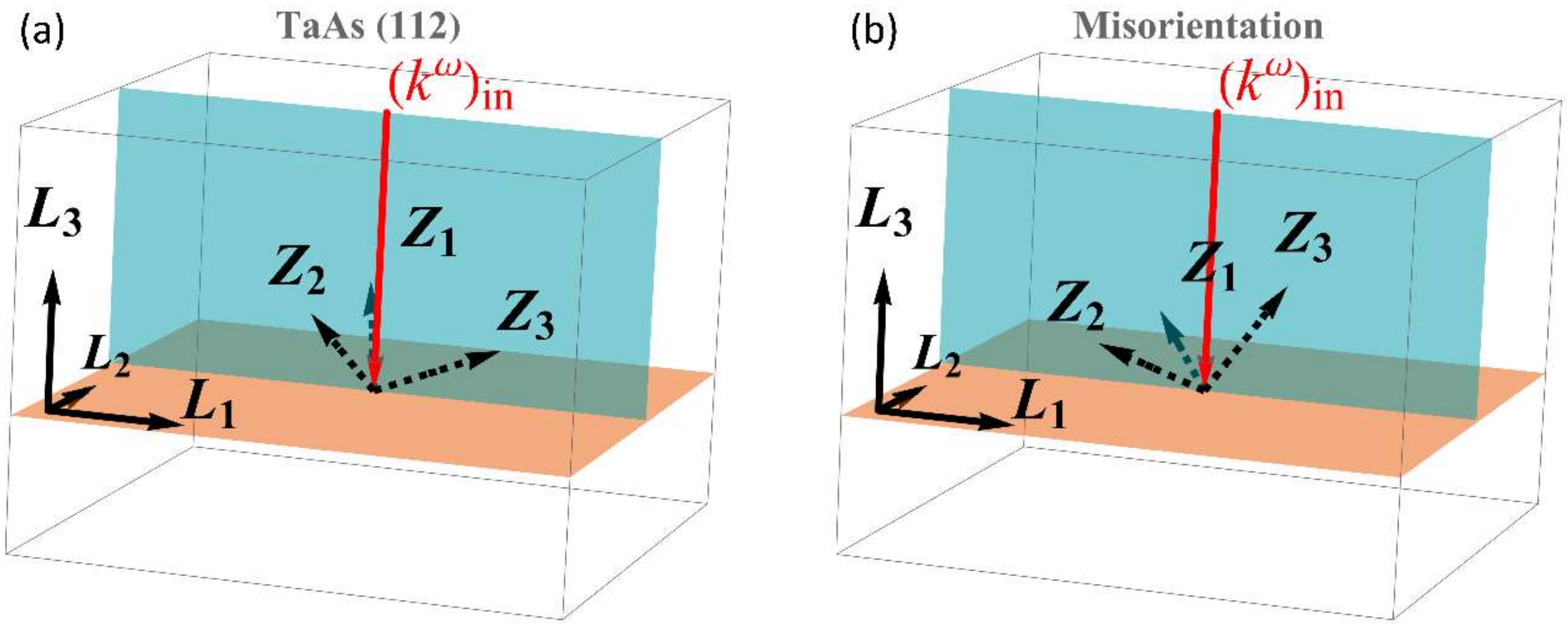

Fig S2. Relation between crystal physics axes $(Z_1, Z_2, Z_3)$ and lab coordinates $(L_1, L_2, L_3)$. Blue and orange planes are the plane of incidence and surface plane, respectively. **a** TaAs(112) with [1 -1 0] parallel to $L_2$. **b** Misorientation with [1 1 -1], [1 -1 0] and [1 1 2] parallel to $L_1$, $L_2$ and $L_3$. $(k^{\omega})_{in}$ is the incident wave vector.

## 3. **Relations between crystal physics axis and crystallographic axis**

Table S1. Relation between crystal physics system (ZCS) and crystallographic system (CCS)

| Lattice system | Optical Class | $\boldsymbol{Z}_3$ | $\boldsymbol{Z}_2$ | $\boldsymbol{Z}_1$ |
|---|---|---|---|---|
| Triclinic | Biaxial | $\boldsymbol{Z}_3 \parallel c$ | $\boldsymbol{Z}_2 \perp ac$ plane (010) | $\boldsymbol{Z}_2 \times \boldsymbol{Z}_3$ |
| Monoclinic | Biaxial | $\boldsymbol{Z}_3 \parallel c$ | $\boldsymbol{Z}_2 \parallel b$ [010] | $\boldsymbol{Z}_2 \times \boldsymbol{Z}_3$ |
| Orthorhombic | Biaxial | $\boldsymbol{Z}_3 \parallel c$ | $\boldsymbol{Z}_2 \parallel b$ | $\boldsymbol{Z}_1 \parallel a$ |
| Trigonal | Uniaxial | $\boldsymbol{Z}_3 \parallel c$ | $\boldsymbol{Z}_3 \times \boldsymbol{Z}_1$ | $\boldsymbol{Z}_1 \parallel a$ [100] |
| Tetragonal | Uniaxial | $\boldsymbol{Z}_3 \parallel c$ | $\boldsymbol{Z}_2 \parallel b$ | $\boldsymbol{Z}_1 \parallel a$ |
| Hexagonal | Uniaxial | $\boldsymbol{Z}_3 \parallel c$ | $\boldsymbol{Z}_2 \parallel b$ | $\boldsymbol{Z}_1 \parallel a$ |
| Cubic | Isotropic | $\boldsymbol{Z}_3 \parallel c$ | $\boldsymbol{Z}_2 \parallel b$ | $\boldsymbol{Z}_1 \parallel a$ |

## 4. Summary of Variables used in #SHAARP

Table S2. Summary of symbols used in #SHAARP.

| Symbols | Explanation | Symbols | Explanation |
|---|---|---|---|
| $(E_{L_1}^{T,e,\omega}, E_{L_2}^{T,e,\omega}, E_{L_3}^{T,e,\omega})$ | Components of the unit vector of the extraordinary electric field at $\omega$ frequency in the $(L_1, L_2, L_3)$ coordinate | $(L_1, L_2, L_3)$ | Orthogonal lab coordinate system (LCS) where experiments are performed. |
| $(E_{L_1}^{T,o,\omega}, E_{L_2}^{T,o,\omega}, E_{L_3}^{T,o,\omega})$ | Components along $L_1$, $L_2$, $L_3$ the direction of the ordinary electric field at $\omega$ frequency. | $(Z_1, Z_2, Z_3)$ | Orthogonal crystal physics coordinate system (ZCS) where materials properties are described. |
| $(E_{L_1}^{T,e,2\omega}, E_{L_2}^{T,e,2\omega}, E_{L_3}^{T,e,2\omega})$ | Components along $L_1$, $L_2$, $L_3$ the direction of the extraordinary electric field at $2\omega$ frequency. | $(a, b, c)$ | Non-orthogonal crystallographic coordinate system (CCS) where lattice parameters are defined. |
| $(E_{L_1}^{T,o,2\omega}, E_{L_2}^{T,o,2\omega}, E_{L_3}^{T,o,2\omega})$ | Components along $L_1$, $L_2$, $L_3$ the direction of the ordinary electric field at $2\omega$ frequency. | $(Z_1^{Principal}, Z_2^{Principal}, Z_3^{Principal})$ | Orthogonal principal coordinate system (PCS) where dielectric/refractive index tensors are diagonalized. |
| $(\tilde{n}_1^{\omega}, \tilde{n}_2^{\omega}, \tilde{n}_3^{\omega})$ | Complex refractive indices in the principal coordinate system at $\omega$. | $(\varepsilon_{L_1L_1}^{\omega}, \varepsilon_{L_1L_2}^{\omega}, \varepsilon_{L_1L_3}^{\omega}, \varepsilon_{L_2L_2}^{\omega}, \varepsilon_{L_2L_3}^{\omega}, \varepsilon_{L_3L_3}^{\omega})$ | Dielectric tensor components at $\omega$ frequency are described in the lab coordinate system |

Table S2. (continued).

| Symbols | Explanation | Symbols | Explanation |
|---|---|---|---|
| $(\tilde{n}_1^{2\omega}, \tilde{n}_2^{2\omega}, \tilde{n}_3^{2\omega})$ | Complex refractive indices in the principal coordinate system at $2\omega$. | $(\varepsilon_{L_1L_1}^{2\omega}, \varepsilon_{L_1L_2}^{2\omega}, \varepsilon_{L_1L_3}^{2\omega}, \varepsilon_{L_2L_2}^{2\omega}, \varepsilon_{L_2L_3}^{2\omega}, \varepsilon_{L_3L_3}^{2\omega})$ | Dielectric tensor components at $2\omega$ frequency are described in the lab coordinate system. |
| $(P_{L_1}^{T,ee,2\omega}, P_{L_2}^{T,ee,2\omega}, P_{L_3}^{T,ee,2\omega})$ | Nonlinear polarization is generated by two extraordinary waves. | $(C_{L_1}^{T,ee,2\omega}, C_{L_2}^{T,ee,2\omega}, C_{L_3}^{T,ee,2\omega})$ | Electric fields radiated by nonlinear polarization $(P_{L_1}^{T,ee,2\omega}, P_{L_2}^{T,ee,2\omega}, P_{L_3}^{T,ee,2\omega})$ |
| $(P_{L_1}^{T,oo,2\omega}, P_{L_2}^{T,oo,2\omega}, P_{L_3}^{T,oo,2\omega})$ | Nonlinear polarization is generated by two ordinary waves. | $(C_{L_1}^{T,oo,2\omega}, C_{L_2}^{T,oo,2\omega}, C_{L_3}^{T,oo,2\omega})$ | Electric fields radiated by nonlinear polarization $(P_{L_1}^{T,oo,2\omega}, P_{L_2}^{T,oo,2\omega}, P_{L_3}^{T,oo,2\omega})$ |
| $(P_{L_1}^{T,eo,2\omega}, P_{L_2}^{T,eo,2\omega}, P_{L_3}^{T,eo,2\omega})$ | Nonlinear polarization is generated by extraordinary and ordinary waves. | $(C_{L_1}^{T,eo,2\omega}, C_{L_2}^{T,eo,2\omega}, C_{L_3}^{T,eo,2\omega})$ | Electric fields radiated by nonlinear polarization $(P_{L_1}^{T,eo,2\omega}, P_{L_2}^{T,eo,2\omega}, P_{L_3}^{T,eo,2\omega})$ |
| $(\theta^i, \theta^R, \theta^{T,e,\omega}, \theta^{T,o,\omega}, \theta^{T,e,2\omega}, \theta^{T,o,2\omega})$ | Incident angle; reflective angle; refractive angle for the extraordinary, ordinary wave at $\omega$; refractive angle for the extraordinary, ordinary wave at $2\omega$ | $(k^i, k^R, k^{T,e,\omega}, k^{T,o,\omega}, k^{T,e,2\omega}, k^{T,o,2\omega}, k^{T,ee,2\omega}, k^{T,oo,2\omega}, k^{T,eo,2\omega})$ | Wavevectors of the incident wave, reflective wave; refractive waves for the extraordinary, ordinary wave at $\omega$; refractive waves for the extraordinary, ordinary wave at $2\omega$; refractive waves for three inhomogeneous wave at $2\omega$ |

## 5. Full Analytical SHG Polarimetry Expressions for GaAs (111)

In this section, the full analytical expressions of GaAs(111) presented in the main text are described in full details. The symbols used in this section are summarized in the **Table. S2**. Due to the isotropic symmetry of GaAs, the refractive indices remain constant regardless of the directions of electric fields, or $\tilde{n}_1^{\omega} = \tilde{n}_2^{\omega} = \tilde{n}_3^{\omega} = \tilde{n}^{\omega}$ and $\tilde{n}_1^{2\omega} = \tilde{n}_2^{2\omega} = \tilde{n}_3^{2\omega} = \tilde{n}^{2\omega}$. Similarly, the angles of refraction for both ordinary and extraordinary waves are the same, which can be written as $\theta^{T,\omega} = \theta^{T,e,\omega} = \theta^{T,o,\omega}$ and $\theta^{T,2\omega} = \theta^{T,e,2\omega} = \theta^{T,o,2\omega}$. In this case, the extraordinary wave is taken as the TM wave and the ordinary wave is taken as the TE wave, where the PoI is the $L_1 - L_3$ plane. Therefore, the unit vectors of electric fields can be represented as $\left(E_{L_1}^{T,e,2\omega}, E_{L_2}^{T,e,2\omega}, E_{L_3}^{T,e,2\omega}\right) = (\cos\theta^{T,2\omega}, 0, \sin\theta^{T,2\omega})$ , $\left(E_{L_1}^{T,o,2\omega}, E_{L_2}^{T,o,2\omega}, E_{L_3}^{T,o,2\omega}\right) = (0,1,0)$ , $\left(E_{L_1}^{T,e,\omega}, E_{L_2}^{T,e,\omega}, E_{L_3}^{T,e,\omega}\right) = (\cos\theta^{T,\omega}, 0, \sin\theta^{T,\omega})$ , and $\left(E_{L_1}^{T,o,\omega}, E_{L_2}^{T,o,\omega}, E_{L_3}^{T,o,\omega}\right) = (0,1,0)$ . Using #SHAARP, the full analytical expressions are calculated and simplified as shown below.

$$E_p^{2\omega} = \frac{\left(C_{L_1}^{T,oo,2\omega} + C_{L_1}^{T,ee,2\omega} + C_{L_1}^{T,eo,2\omega}\right)\left(E_{L_1}^{T,e,2\omega}\tilde{n}^{2\omega}\cos\theta^{T,2\omega} - E_{L_1}^{T,e,2\omega}\tilde{n}^{\omega}\cos\theta^{T,\omega} + E_{L_3}^{T,e,2\omega}\tilde{n}^{2\omega}\sin\theta^{T,2\omega}\right) - \left(C_{L_3}^{T,oo,2\omega} + C_{L_3}^{T,ee,2\omega} + C_{L_3}^{T,eo,2\omega}\right)E_{L_1}^{T,e,2\omega}\tilde{n}^{\omega}\sin\theta^{T,\omega}}{E_{L_1}^{T,e,2\omega} + \tilde{n}^{2\omega}\cos\theta^{i}\left(E_{L_1}^{T,e,2\omega}\cos\theta^{T,2\omega} + E_{L_3}^{T,e,2\omega}\sin\theta^{T,2\omega}\right)} \tag{S9}$$

$$E_s^{2\omega} = \frac{\left(C_{L_2}^{T,oo,2\omega} + C_{L_2}^{T,ee,2\omega} + C_{L_2}^{T,eo,2\omega}\right)\left(\tilde{n}^{2\omega}\cos\theta^{T,2\omega} - \tilde{n}^{\omega}\cos\theta^{T,\omega}\right)}{\tilde{n}^{2\omega}\left(\cos\theta^{T,2\omega} + \cos\theta^{i}\right)} \tag{S10}$$

$$C_{L_1}^{T,ee,2\omega} = \frac{-\left(\tilde{n}^{2\omega}\right)^2 P_{L_1}^{T,ee,2\omega}\mu_0 - (\tilde{n}^{\omega})^2 P_{L_3}^{T,ee,2\omega}\omega^2\cos\theta^{T,\omega}\sin\theta^{T,\omega} + (\tilde{n}^{\omega})^2 P_{L_1}^{T,ee,2\omega}\omega^2\sin^2\theta^{T,\omega}}{(\tilde{n}^{2\omega})^2((\tilde{n}^{2\omega})^2\mu_0 - (\tilde{n}^{\omega})^2\omega^2)\varepsilon_0} \tag{S11}$$

$$C_{L_2}^{T,ee,2\omega} = -\frac{P_{L_2}^{T,ee,2\omega}\mu_0}{(\tilde{n}^{2\omega})^2\mu_0\varepsilon_0 - (\tilde{n}^{\omega})^2\omega^2\varepsilon_0} \tag{S12}$$

$$C_{L_3}^{T,ee,2\omega} = \frac{-(\tilde{n}^{2\omega})^2 P_{L_3}^{T,ee,2\omega}\mu_0 + (\tilde{n}^{\omega})^2 P_{L_3}^{T,ee,2\omega}\omega^2\cos^2\theta^{T,\omega} - (\tilde{n}^{\omega})^2 P_{L_1}^{T,ee,2\omega}\omega^2\cos\theta^{T,\omega}\sin\theta^{T,\omega}}{(\tilde{n}^{2\omega})^2((\tilde{n}^{2\omega})^2\mu_0 - (\tilde{n}^{\omega})^2\omega^2)\varepsilon_0} \tag{S13}$$

$$C_{L_1}^{T,oo,2\omega} = \frac{-\left(\tilde{n}^{2\omega}\right)^2 P_{L_1}^{T,oo,2\omega}\mu_0 - (\tilde{n}^{\omega})^2 P_{L_3}^{T,oo,2\omega}\omega^2\cos\theta^{T,\omega}\sin\theta^{T,\omega} + (\tilde{n}^{\omega})^2 P_{L_1}^{T,oo,2\omega}\omega^2\sin^2\theta^{T,\omega}}{(\tilde{n}^{2\omega})^2((\tilde{n}^{2\omega})^2\mu_0 - (\tilde{n}^{\omega})^2\omega^2)\varepsilon_0} \tag{S14}$$

$$C_{L_2}^{T,oo,2\omega} = -\frac{P_{L_2}^{T,oo,2\omega}\mu_0}{(\tilde{n}^{2\omega})^2\mu_0\varepsilon_0 - (\tilde{n}^{\omega})^2\omega^2\varepsilon_0} \tag{S15}$$

$$C_{L_3}^{T,oo,2\omega} = \frac{-(\tilde{n}^{2\omega})^2 P_{L_3}^{T,\mathrm{oo},2\omega}\mu_0 + (\tilde{n}^{\omega})^2 P_{L_3}^{T,\mathrm{oo},2\omega}\omega^2\cos^2\theta^{T,\omega} - (\tilde{n}^{\omega})^2 P_{L_1}^{T,\mathrm{oo},2\omega}\omega^2\cos\theta^{T,\omega}\sin\theta^{T,\omega}}{(\tilde{n}^{2\omega})^2((\tilde{n}^{2\omega})^2\mu_0 - (\tilde{n}^{\omega})^2\omega^2)\varepsilon_0} \tag{S16}$$

$$C_{L_1}^{T,eo,2\omega} = \frac{-(\tilde{n}^{2\omega})^2 P_{L_1}^{T,\mathrm{eo},2\omega}\mu_0 - (\tilde{n}^{\omega})^2 P_{L_3}^{T,\mathrm{eo},2\omega}\omega^2\cos\theta^{T,\omega}\sin\theta^{T,\omega} + (\tilde{n}^{\omega})^2 P_{L_1}^{T,\mathrm{eo},2\omega}\omega^2\sin^2\theta^{T,\omega}}{(\tilde{n}^{2\omega})^2((\tilde{n}^{2\omega})^2\mu_0 - (\tilde{n}^{\omega})^2\omega^2)\varepsilon_0} \tag{S17}$$

$$C_{L_2}^{T,eo,2\omega} = -\frac{P_{L_2}^{T,\mathrm{eo},2\omega}\mu_0}{(\tilde{n}^{2\omega})^2\mu_0\varepsilon_0 - (\tilde{n}^{\omega})^2\omega^2\varepsilon_0} \tag{S18}$$

$$C_{L_3}^{T,eo,2\omega} = \frac{-(\tilde{n}^{2\omega})^2 P_{L_3}^{T,\mathrm{eo},2\omega}\mu_0 + (\tilde{n}^{\omega})^2 P_{L_3}^{T,\mathrm{eo},2\omega}\omega^2\cos^2\theta^{T,\omega} - (\tilde{n}^{\omega})^2 P_{L_1}^{T,\mathrm{eo},2\omega}\omega^2\cos\theta^{T,\omega}\sin\theta^{T,\omega}}{(\tilde{n}^{2\omega})^2((\tilde{n}^{2\omega})^2\mu_0 - (\tilde{n}^{\omega})^2\omega^2)\varepsilon_0} \tag{S19}$$

$$P_{L_1}^{T,\mathrm{ee},2\omega} = E_{L_1}^{e,\omega}(1.63299E_{L_2}^{e,\omega} - 1.15470E_{L_3}^{e,\omega})d_{14}\varepsilon_0 \tag{S20}$$

$$P_{L_2}^{T,\mathrm{ee},2\omega} = (0.81650(E_{L_1}^{e,\omega})^2 + E_{L_2}^{e,\omega}(-0.81650E_{L_2}^{e,\omega} - 1.15470))d_{14}\varepsilon_0 \tag{S21}$$

$$P_{L_3}^{T,\mathrm{ee},2\omega} = (-0.57735(E_{L_1}^{e,\omega})^2 - 0.57735(E_{L_2}^{e,\omega})^2 + 1.15470(E_{L_3}^{e,\omega})^2)d_{14}\varepsilon_0 \tag{S22}$$

$$P_{L_1}^{T,\mathrm{oo},2\omega} = E_{L_1}^{o,\omega}(1.63299E_{L_2}^{o,\omega} - 1.15470E_{L_3}^{o,\omega})d_{14}\varepsilon_0 \tag{S23}$$

$$P_{L_2}^{T,\mathrm{oo},2\omega} = (0.81650(E_{L_1}^{o,\omega})^2 + E_{L_2}^{o,\omega}(-0.81650E_{L_2}^{o,\omega} - 1.15470E_{L_3}^{o,\omega}))d_{14}\varepsilon_0 \tag{S24}$$

$$P_{L_3}^{T,\mathrm{oo},2\omega} = (-0.57735(E_{L_1}^{o,\omega})^2 - 0.57735(E_{L_2}^{o,\omega})^2 + 1.15470(E_{L_3}^{o,\omega})^2)d_{14}\varepsilon_0 \tag{S25}$$

$$P_{L_1}^{T,\mathrm{eo},2\omega} = (1.63299E_{L_1}^{o,\omega}E_{L_2}^{e,\omega} + 1.63299E_{L_2}^{o,\omega} - 1.15470E_{L_1}^{o,\omega}E_{L_3}^{e,\omega} - 1.15470E_{L_1}^{e,\omega}E_{L_3}^{o,\omega})d_{14}\varepsilon_0 \tag{S26}$$

$$P_{L_2}^{T,\mathrm{eo},2\omega} = (1.63299E_{L_1}^{e,\omega}E_{L_1}^{o,\omega} - 1.63299E_{L_2}^{o,\omega} - 1.15470E_{L_2}^{o,\omega}E_{L_3}^{e,\omega} - 1.15470E_{L_2}^{e,\omega}E_{L_3}^{o,\omega})d_{14}\varepsilon_0 \tag{S27}$$

$$P_{L_3}^{T,\mathrm{eo},2\omega} = (-1.15470E_{L_1}^{o,\omega} - 1.15470E_{L_2}^{e,\omega}E_{L_2}^{o,\omega} + 2.30940E_{L_3}^{e,\omega}E_{L_3}^{o,\omega})d_{14}\varepsilon_0 \tag{S28}$$

$$E_{L_1}^{e,\omega} = \frac{2E_{L_1}^{T,e,\omega}\cos\varphi\cos\ {}^{i}}{E_{L_1}^{T,e,\omega} + E_{L_1}^{T,e,\omega}\tilde{n}^{\omega}\cos\theta^{T,\omega}\cos\theta^{i} + E_{L_3}^{T,e,\omega}\tilde{n}^{\omega}\cos\theta^{i}\sin\theta^{T,\omega}} \tag{S29}$$

$$E_{L_2}^{e,\omega} = 0 \tag{S30}$$

$$E_{L_3}^{e,\omega} = \frac{2E_{L_3}^{T,e,\omega}\cos\varphi\cos\theta^{i}}{E_{L_1}^{T,e,\omega} + E_{L_1}^{T,e,\omega}\tilde{n}^{\omega}\cos\theta^{T,\omega}\cos\theta^{i} + E_{L_3}^{T,e,\omega}\tilde{n}^{\omega}\cos\theta^{i}\sin\theta^{T,\omega}} \tag{S31}$$

$$E_{L_1}^{o,\omega} = 0 \tag{S32}$$

$$E_{L_2}^{o,\omega} = \frac{2\cos\theta_{\mathrm{in}}\sin\varphi}{\tilde{n}^{\omega}\cos\theta^{T,\omega} + \cos\theta^{i}} \tag{S33}$$

$$E_{L_3}^{o,\omega} = 0 \tag{S34}$$

**Reference**

1. Wu, L. *et al.* Giant anisotropic nonlinear optical response in transition metal monopnictide Weyl semimetals. *Nature Physics* **13**, 350–355 (2017).
2. Patankar, S. *et al.* Resonance-enhanced optical nonlinearity in the Weyl semimetal TaAs. *Phys. Rev. B* **98**, 165113 (2018).
3. Shoji, I., Kondo, T., Kitamoto, A., Shirane, M. & Ito, R. Absolute scale of second-order nonlinear-optical coefficients. *J. Opt. Soc. Am. B, JOSAB* **14**, 2268–2294 (1997).
4. Nikogosyan, D. N. *Nonlinear Optical Crystals: A Complete Survey*. (Springer-Verlag, 2005). doi:10.1007/b138685.